\documentclass[twocolumn]{aastex62}
\received{MMM D, YYYY}
\revised{MMM D, YYYY}
\accepted{MMM D, YYYY}
\submitjournal{ApJ}
\usepackage{threeparttable}
\usepackage{soul}
\usepackage{graphicx}
\shorttitle{DI and Chromospheric Analysis of PW And}
\shortauthors{Bahar et al.}

\begin{document}

\title{First Chromospheric Activity and Doppler Imaging Study of PW And Using a New Doppler Imaging Code: SpotDIPy}

\correspondingauthor{E. Bahar}
\email{enbahar@ankara.edu.tr}
\author[0000-0001-9119-2316]{Engin Bahar}
\affiliation{Department of Astronomy and Space Sciences, Faculty of Science, Ankara University, Be\c{s}evler 06100, Ankara, T\"{u}rkiye}

\author[0000-0002-8961-277X]{Hakan V. \c{S}enavc{\i}}
\affiliation{Department of Astronomy and Space Sciences, Faculty of Science, Ankara University, Be\c{s}evler 06100, Ankara, T\"{u}rkiye}

\author[0000-0001-6163-0653]{Emre I\c{s}{\i}k}
\affiliation{Max-Planck-Institut f\"{u}r Sonnensystemforschung, Justus-von-Liebig-Weg 3, 37077 G\"{o}ttingen, Germany}
\affiliation{Department of Computer Science, Turkish-German University, Beykoz 34820, Istanbul, Türkiye}

\author[0000-0003-3547-3783]{Gaitee A.J. Hussain}
\affiliation{European Space Agency (ESA), European Space Research and Technology Centre (ESTEC), Keplerlaan 1, 2201 AZ Noordwijk, The Netherlands}

\author[0000-0003-3061-4591]{Oleg Kochukhov}
\affiliation{Department of Physics and Astronomy, Uppsala University, Box 516, SE-75120 Uppsala, Sweden}

\author[0000-0002-7779-238X]{David Montes}
\affiliation{Departamento de F{\'i}sica de la Tierra y Astrof{\'i}sica \& IPARCOS-UCM (Instituto de F\'{i}sica de Part\'{i}culas y del Cosmos de la UCM), Facultad de Ciencias F{\'i}sicas, Universidad Complutense de Madrid, E-28040 Madrid, Spain}

\author[0000-0001-6580-9378]{Yue Xiang}
\affiliation{Yunnan Observatories, Chinese Academy of Sciences, Kunming 650216, People’s Republic of China}
\affiliation{Key Laboratory for the Structure and Evolution of Celestial Objects, Chinese Academy of Sciences, Kunming 650216, People’s Republic of China}



\begin{abstract}

Measuring coverage of dark spots on cool stars is important in understanding how stellar magnetic activity scales with the rotation rate and convection zone depth. In this respect, it is crucial to infer surface magnetic patterns on G and K stars, to reveal solar-like stellar dynamos in action. Molecular bands serve as invaluable indicators of cool spots on the surfaces of stars, as they play a crucial role in enabling accurate assessments of the extent of spot coverage across the stellar surface. Therefore, more reliable surface images can be obtained considering the inversion of atomic lines with molecular bands. In this context, we simultaneously carry out Doppler imaging (DI) using atomic lines as well as Titanium Oxide (TiO) band profiles of PW And (K2~V) and also investigate chromospheric activity indicators for the first time in the literature, using the high-resolution spectra. The surface spot distribution obtained from the inversion process represents both atomic line and TiO-band profiles quite accurately. The chromospheric emission is also correlated with photospheric spot coverage, except during a possible flare event during the observations. We detect frequent flare activity, using TESS photometry. We also introduce a new open-source, Python-based DI code {\tt SpotDIPy} that allows performing surface reconstructions of single stars using the maximum entropy method. We test the code by comparing surface reconstruction simulations with the extensively used {\tt DoTS} code. We show that the surface brightness distribution maps reconstructed via both codes using the same simulated data are consistent with each other.

\end{abstract}

\keywords{stars: activity, stars: imaging -- starspots }

\section{Introduction}
\label{sec:intro}

PW And is a rapidly rotating pre-main-sequence K2~V-type star, which is a member of the AB Doradus moving group. Its high Li~{\sc i} abundance also confirms its membership to the young Local Association \citep{Montes2001a, Montes2001b, Lopez2003, Montes2004}. Its projected rotational velocity, $v\sin i$, was determined by several investigators in the literature in the range from 21.5 km~s$^{-1}$ \citep{Griffin1992} to 23.9 km~s$^{-1}$ \citep{Strassmeier2006}. A detailed activity investigation of PW And was performed by \citet{Lopez2003}, who analysed photospheric Ca~{\sc i} and Fe~{\sc i} line profiles using CCF bisector analysis \citep[see][for more details]{Dempsey1992} and found that the CCF bisectors change with a period similar to the photometric period of PW And, suggesting the presence of cool spots. Using the spectral subtraction technique \citep{Montes2000}, they also measured the equivalent width variation of chromospheric activity indicator lines from Ca~{\sc ii}~H\&K to Ca~{\sc ii}~IRT lines along with the rotational phase and found that the  chromospheric regions appear to be associated with the photospheric features obtained using CCF bisector analysis. In addition, they detected two flares in 2001 and 2002, confirming the strong magnetic activity of the star. \citet{Lopez2010} and \citet{Lehtinen2016} determined emission index as $log R^{'}_{HK}$ -3.85 and -4.217, respectively, revealing the highly active chromosphere. The first DI of PW And was obtained by \citet{Strassmeier2006}, using high-resolution CFHT spectroscopy. They found that cool spots are located within an equatorial band up to $\pm40^{\circ}$ of the stellar equator, with temperature contrast relative to the immaculate photosphere of up to $\Delta T = 1200$~K. They also estimated a set of astrophysical parameters \citep[see Table 1 of][for more details]{Strassmeier2006}, including the rotation period as 1.76159 days, using high-precision photometry. Another DI analysis was carried out by \citet{Gu2010}, who used high-resolution spectra obtained at the Xinglong Station of National Astronomical Observatories (NAOC) and Bohyunsan Astronomical Observatory (BOAO). Their spectra were separated into two subsets spanning approximately one month. Unlike the Doppler images presented by \citet{Strassmeier2006}, their resulting spot distributions were rather concentrated within intermediate to high latitudes, while weaker low-latitude spots were also visible. They also reported that there is no notable position variation of spots and concluded that intermediate- to high-latitude spots have lifetimes longer than a month. Based on the high-resolution spectroscopic observations carried out at the RTT-150 telescope of T\"{U}B\.{I}TAK National Observatory, the most recent DI was performed by \citet{Kolbin2017}. They found that spots were located around latitude $40^{\circ}$ and claimed that the resultant map is similar to that obtained by \citet{Strassmeier2006}, where spots spread between $0^{\circ}$-$40^{\circ}$ with a tendency toward $30^{\circ}$.

The latitudinal distributions of spots on PW And that were inferred in previous studies are substantially different from each other. Possible underlying reasons are the image reconstruction process and intrinsic variations of spot distribution on the star between different epochs. New observations carried out at different epochs are thus needed to improve the basic statistics of the latitudinal spot distribution. Another reason for reexamining PW And, apart from previous DI studies based on atomic lines, is to simultaneously confirm the spot distribution and contrast using TiO-band at 7055~$\mbox{\AA}$ in this research.

In this study, we conducted a simultaneous analysis of both atomic line and TiO-band profiles with the DI technique on PW And, using three distinct sets of observations spanning different time intervals. Concurrently, we investigated chromospheric activity within the same dataset and compared the observed activity trends. Additionally, an accurate rotation period of PW And was calculated using the Lomb-Scargle periodogram and $v\sin i$ was determined on a 2D grid search. We also introduced a new open-source and user friendly Python-based DI code {\tt SpotDIPy} and tested via {\tt DoTS} \citep{Cameron1992}. 

\section{Observations and data reduction} \label{sec:obs}

We obtained a high-resolution spectral time-series of PW And, using the HERMES\footnote{Based on observations obtained with the HERMES spectrograph mounted on the 1.2 m Mercator Telescope at the Spanish Observatorio del Roque de los Muchachos of the Instituto de Astrof{\'i}sica de Canarias.} spectrograph \citep{Raskin2011} attached to the 1.2-~m Mercator telescope at the Roque de los Muchachos Observatory (La Palma, Spain), between 14-19 December in 2015 (hereafter Set-1), 25-29 September in 2018 (hereafter Set-2) and 14-19 December in 2018 (hereafter Set-3). Consequently, Set-1, Set-2 and Set-3 data of PW And obtained in this study cover approximately 2.9, 2.4 and 2.3 rotational cycles, respectively.

The average spectral resolution is $R$=85\,000 with a wavelength coverage between 3780 and 9007~$\mbox{\AA}$. The data were acquired using exposure times between 1200 and 1800 seconds, which yielded signal-to-noise ratios (S/N) between 73 and 174. The S/N values of the spectra are given as per pixel around 550 nm. The effective spectral resolution is between 0.045-0.106~$\mbox{\AA}$ in wavelength and 3.5 km/sn in velocity units, while the sampling is 2 pixels per resolution element. The log of HERMES observing run is given in Table~\ref{tab:obslog}. The reduction of the spectra were performed using the automatic pipeline of the spectrograph \citep{Raskin2011}. The pipeline was utilized to execute fundamental echelle data reduction procedures, encompassing tasks such as bias, flat-field and the inter-order background level corrections, as well as the extraction of 1D spectra and subsequent wavelength calibration. The normalization procedure was carried out via a Python code developed by our working group \citep{Senavci2018}.

\begin{figure}
\centering
	\includegraphics[width=\columnwidth]{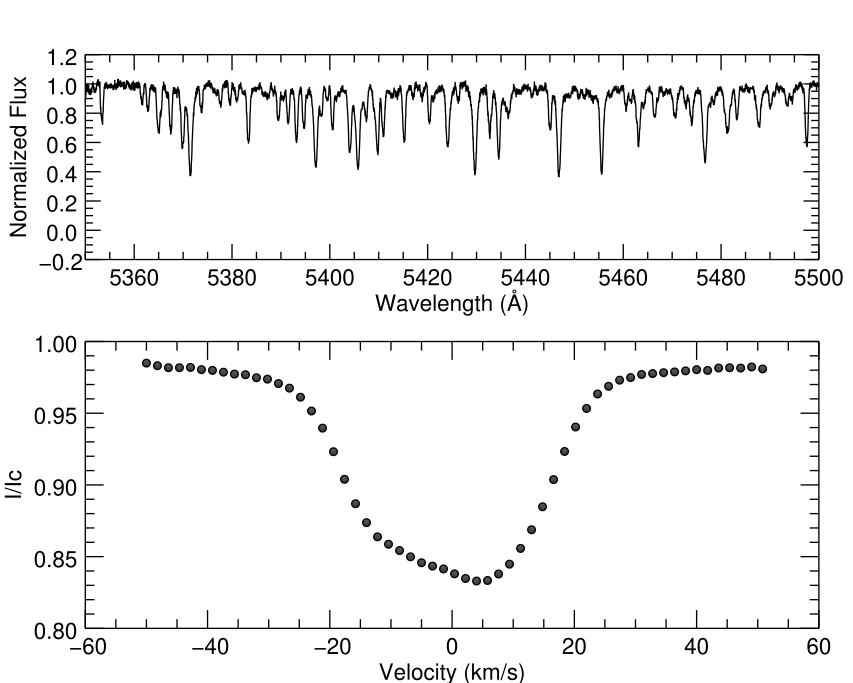}
 	\caption{The input spectrum (top panel) and the resultant LSD profile (bottom panel) of PW And, corresponding to the rotational phase 0.002 of Set-1. Note that only a part of the input spectrum is shown here for better visualization.}
    \label{fig:speclsd}
\end{figure}

We used the multi-line technique, Least Squares Deconvolution (hereafter LSD) \citep{Donati1997b}, to produce a mean photospheric line profile with higher S/N, in turn improving the quality of the resulting spot maps. The line mask, contains information on the line positions and relative strengths, and is required to construct the mask used by LSD. We extracted the line mask from the Vienna Atomic Line Database (VALD) \citep{Kupka1999}. The wavelength regions including lines that are affected by chromospheric heating (i.e. Hydrogen Balmer series, Na~{\sc i} (D$_{\rm 1}$, D$_{\rm 2}$), Ca~{\sc ii} IRT) and strong telluric lines were removed from the list to prevent any artifacts in the LSD profiles. We set the velocity increment to 1.75 km~s$^{-1}$, considering spectral resolution (3.5 km~s$^{-1}$) and its sampling (2) in pixel. We obtained LSD profiles of PW And with S/N in the range between 670 and 850. An example of a part of an input spectrum and the resultant LSD profile are shown in Figure~\ref{fig:speclsd}.

The Transiting Exoplanet Survey Satellite \citep[TESS]{Ricker2015} photometry of PW And enabled us to determine the rotation period with high precision. TESS observed PW And in sectors 17 and 57, corresponding to time spans of October 8 - November 2, 2019 and September 30 - October 29, 2022, respectively. All of the light curve data presented in this paper were obtained from the Mikulski Archive for Space Telescopes (MAST\footnote{https://mast.stsci.edu}) at the Space Telescope Science Institute. The specific observations analyzed can be accessed via \dataset[DOI]{https://doi.org/10.17909/g29a-2t15}. Among the two types of photometric data available, SAP flux is generated by summing all pixel values in a pre-defined aperture as a function of time, while PDCSAP flux is SAP flux from which long term trends have been removed, using so-called Co-trending Basis Vectors (CBVs). Since there is no amplitude difference between SAP and PDCSAP light curve data, we used the latter one to determine the rotation period of PW And.

\begin{table}
	\centering
	\caption{Phase-ordered spectroscopic observation log of PW And.}
	\label{tab:obslog}
	\begin{tabular}{lcccc} 
	\hline
        \hline
	Date & Exp. Time  & BJD$_{Mid}$ & Phase$_{Mid}$ & S/N\\
    	&	(sec.)&	&	&\\
	\hline
        \hline
18.12.2015&1800&57375.45063&0.002&108\\
15.12.2015&1200&57372.32557&0.223&74\\
19.12.2015&1800&57376.30420&0.488&77\\
19.12.2015&1800&57376.32584&0.500&73\\
19.12.2015&1800&57376.47665&0.586&104\\
14.12.2015&1600&57371.33483&0.659&92\\
16.12.2015&1800&57373.31890&0.788&102\\
18.12.2015&1800&57375.30685&0.920&108\\
\hline
25.09.2018&1500&58387.36672&0.066&155\\
25.09.2018&1500&58387.45742&0.117&172\\
27.09.2018&1500&58389.40298&0.225&119\\
27.09.2018&1500&58389.47599&0.266&142\\
29.09.2018&1500&58391.39726&0.360&170\\
29.09.2018&1500&58391.50205&0.420&174\\
26.09.2018&1500&58388.40970&0.659&169\\
27.09.2018&1500&58388.51470&0.719&169\\
\hline
15.12.2018&1600&58468.35045&0.168&79\\
17.12.2018&1600&58470.31374&0.286&154\\
17.12.2018&1600&58470.43395&0.354&161\\
18.12.2018&1200&58470.51817&0.402&135\\
14.12.2018&1600&58467.44046&0.650&105\\
16.12.2018&1600&58469.35515&0.740&111\\
16.12.2018&1600&58469.46048&0.800&126\\
18.12.2018&1200&58471.31222&0.854&140\\
18.12.2018&1200&58471.43170&0.922&144\\
19.12.2018&1200&58471.51305&0.968&129\\

        \hline
        \hline
	\end{tabular}
\end{table}

\section{Analysis}
\subsection{Rotation Period Estimation}

\begin{figure*}
\centering
	\includegraphics[width=42pc,height=21pc]{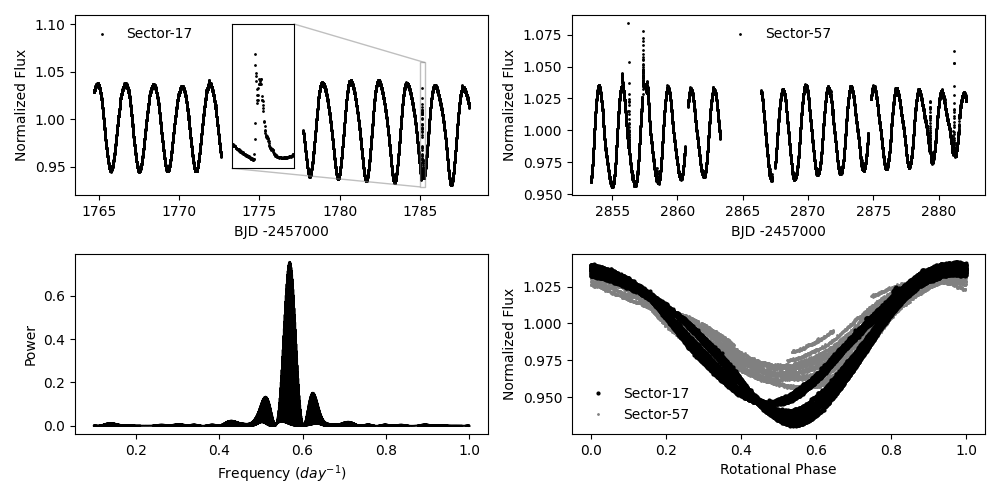}
 	\caption{Top panel: TESS light curves in Sector 17 (left) and in Sector 57 (right). Bottom panel: Lomb-Scargle periodogram (left) and phase-folded light curves (right) using the period determined from the periodogram, for Sector 17 (black circles) and Sector 57 (gray circles).}
    \label{fig:lc_and_period}
\end{figure*}

Although the TESS sectors 17 and 57 are about three years apart, the light curve morphology remains almost the same in both sectors as a dominant sinusoidal variation. This is shown in Figure~\ref{fig:lc_and_period} (upper panels). To investigate periodicities owing to the rotational modulation by spots, we computed the Lomb-Scargle periodogram, using the light curves from both sectors, after removing outliers and flares. The periodogram shows a clear peak (Figure~\ref{fig:lc_and_period}, lower left panel), corresponding to 0.569280 $days^{-1}$ in the frequency domain and 1.756604 $\pm$0.000015 days in the period domain. The period and its associated uncertainty were determined by fitting the peak using a Gaussian function through the least-squares method, along with the application of the bootstrap method. The resultant period is close to 1.76159 days obtained by \citet{Strassmeier2006}. The phase-folded light curves in Figure~\ref{fig:lc_and_period} show that the Sector 17 data has a higher amplitude compared to Sector 57, while both sector data show similar modulations. All the photometric and spectroscopic data used in analysis are phased using the following equation:

\begin{equation} \label{eq:ephemeris}
{\rm BJD} = 2453200.00 + 1^{\rm d}.756604 \times E.
\end{equation}

\subsection{Stellar Parameters Estimation}

\begin{figure*}
\centering
	\includegraphics[width=\textwidth]{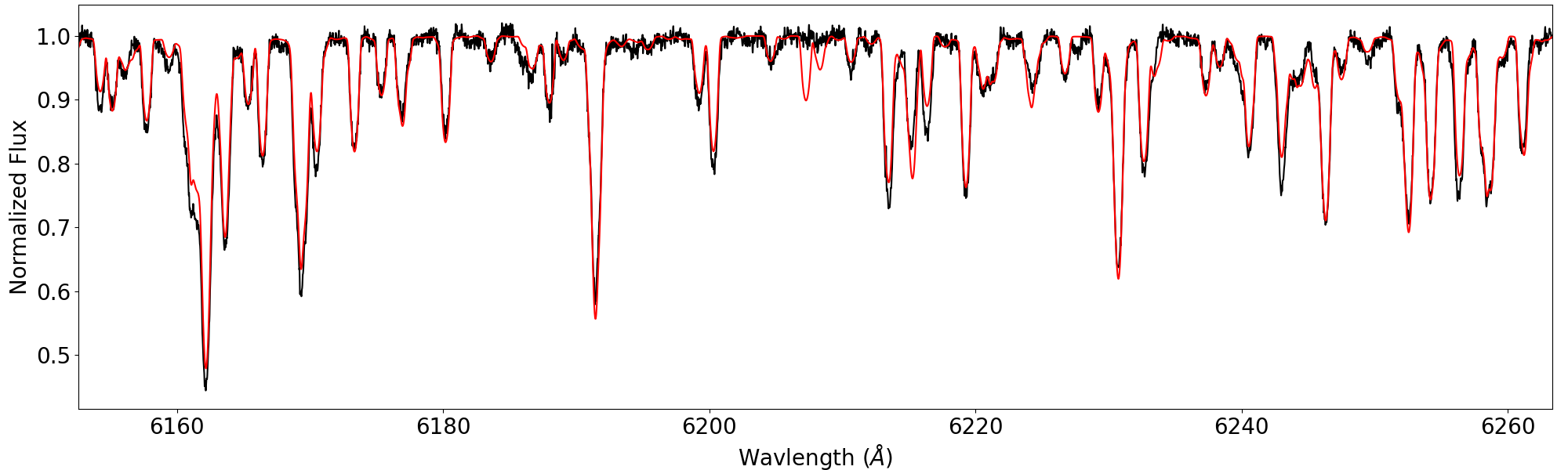}
 	\caption{A comparison of the observed spectrum of PW And (black) with the best fit model (red) obtained through spectral synthesis fitting method.}
    \label{fig:specfit}
\end{figure*}

We obtained the stellar parameters of PW And using the spectral synthesis fitting method. Synthetic spectra were produced using the SPECTRUM \citep{Gray1994} code with MARCS \citep{Gustafsson2008} atmosphere models and the VALD line list. The fitting was performed on all Fe I and Fe II lines in the wavelength range of 5400 - 6800~$\mbox{\AA}$ of the stellar spectrum. This region was selected because it contains relatively few telluric lines and the continuum can be more easily detected. All of these operations were carried out through iSpec \citep{Blanco2014, Blanco2019}. The obtained parameters and errors were calculated using the bootstrap method. The resultant parameters are given in Table~\ref{tab:sysparams}, which is consistent with the parameters obtained by \citet{Folsom2016}, recently. A comparison of the synthetic spectrum and the observed spectrum is shown in Figure~\ref{fig:specfit}.

\subsection{Doppler Imaging Code: {\tt SpotDIPy}}

To reconstruct the distribution of starspots, we developed a DI code, {\tt SpotDIPy}, written in Python programming language. The code generates synthetic line profiles ($R_{calc}$) by integrating local line profiles across the entire visible stellar surface at each rotational phase, $\phi$. Each surface element on the stellar surface consist of two line profiles with different temperatures, one representing the photosphere and the other representing the spots. This is a basic assumption for the DI method based on the two-temperature model with the resulting distribution in terms of spot filling factors, $f_s$, defined as the fraction of the surface element area occupied by spots. During the integration process, local line profiles are shifted and scaled with respect to the position of the surface element in the velocity space and the projected areas, respectively. The local line profiles also are scaled with continuum flux ratio calculated from blackbody assumption based on their corresponding temperatures and central wavelengths. Limb- and gravity-darkening effects are taken into account during reconstruction. Limb-darkening coefficients are calculated by a Python package, {\tt ExoTiC-LD} \citep{grant2022}. The intensity factors due to the gravity-darkening effect are determined, considering the study by \citet{Espinosa2011}. {\tt SpotDIPy} has two surface-grid mode, triangulation (supplied from {\tt PHOEBE2}\footnote[1]{github.com/phoebe-project/phoebe2} Python package \citep{Prsa2016}) and trapezoidal discretization based on trapezoidal elements. 

{\tt SpotDIPy} uses the Maximum Entropy Method (MEM) to solve the ill-posed inverse problem, by taking a regularization function of the form 
\begin{equation} \label{eq:mem}
S(f_s)=-\sum_{i=1}^{n} w_{i} \left [f_{s_i}log\frac{f_{s_i}}{m}+(1-f_{s_i})log\frac{(1-f_{s_i})}{(1- m)} \right ],
\end{equation}
where $w_i$ is the area of the $i^{th}$ surface element and $m$ is the default minimum fraction of spot coverage that is set to a very small, positive value. 
To find the best-fitting spot distribution over the visible part of the stellar surface, we use the error function
\begin{equation} \label{eq:error}
E=\sum_{\phi} \sum_{\nu} \left[\frac{R_{calc}(\nu, \phi)-R_{obs}(\nu, \phi)}{\sigma_{\nu, \phi}} \right]^{2}+
\lambda S,
\end{equation}
where $R_{calc}$ and $R_{obs}$ are the calculated and observed mean line profiles (as a function of radial velocity $\nu$) at phase $\phi$ with uncertainty $\sigma$, and $\lambda$ is the Lagrange multiplier \citep[see][for more details of Doppler imaging based on two-temperature model]{Cameron1992}. {\tt SpotDIPy} searches for the best Lagrange multiplier using the method suggested by \citet{Chiang2005}, where minimum $\chi^2$ and the entropy are calculated for each Lagrange multiplier during the optimization process. Then, the Lagrange multiplier that corresponds to the maximum curvature of the function $\chi^2$ of entropy is used during the DI reconstruction. {\tt SpotDIPy} uses L-BFGS-B algorithm \citep[see][for more details]{Byrd1995, Zhu1997} to solve the bound constrained optimization problem, which is available in the {\tt SciPy}\footnote{https://scipy.org/} library.

To test the robustness of {\tt SpotDIPy}, we use {\tt DoTS} as the comparison case, since the latter has been extensively used in the literature for DI purposes. In this context, using {\tt DoTS} code, we generated synthetic line profiles with the S/N value of 500, including five spots with different sizes and locations, to simulate a spotted single star. We also assumed three different axial inclinations as 30$^{\circ}$, 60$^{\circ}$ and 90$^{\circ}$. The synthetic line profiles were generated with a sampling interval of 0.1 in phase. The stellar parameters were adopted as $v\sin i = 40$ ~km~s$^{-1}$, $mass=1$ $M_\odot$ and $period=1.5 ~days$. These profiles were then used as input data for the surface reconstruction process performed using both {\tt DoTS} and {\tt SpotDIPy}. The surface grids for both codes were generated using 90 latitudinal zones that correspond to 180 surface elements along the equator, which is above the lower limit (80 surface elements) for the adopted $v\sin i$ value and the resolution ($R$=150000). Optimization was carried out until the same $\chi^2$ value is obtained in both codes as $\sim$1. Consequently, we calculated the standard deviations of the residuals, differences between models obtained from {\tt DoTS} and {\tt SpotDIPy}, in the order of $4.8\times10^{-4}$, $3.9\times10^{-4}$ and $3.4\times10^{-4}$, for the three different axial inclination cases. The resultant maps generated by {\tt SpotDIPy} are quite similar to those of {\tt DoTS}, as seen in Figures~\ref{fig:comp_dots_sdip30}, ~\ref{fig:comp_dots_sdip60} and ~\ref{fig:comp_dots_sdip90}. Therefore, the test results showed that {\tt SpotDIPy} can perform DI with the same accuracy as {\tt DoTS}, for single stars.

In addition to the automatic calculation of the limb- and gravity-darkening coefficients, {\tt SpotDIPy} also takes into account the effects of macroturbulence and instrumental profile, unlike the DoTS code. {\tt SpotDIPy} applies the macroturbulence broadening using the radial-tangential formulation adapted from SME \citep{Valenti1996}. The instrumental profile is taken into account by convolving a kernel calculated over the given spectral resolution with the local line profiles, which is performed using the {\tt PyAstronomy}\footnote{https://github.com/sczesla/PyAstronomy} \citep{pya2019} Python package.

Molecular bands, such as Titanium Oxide (TiO), play a crucial role in providing precise assessments of spot coverage on the surfaces of stars. The area covered by cool spots as a percentage of the total stellar surface area can be determined through the analysis of molecular bands \citep{Berdyugina2002}. In this context, We developed {\tt SpotDIPy} to reconstruct surface maps by simultaneously utilizing both atomic lines and molecular band profiles. The reconstruction process of TiO-bands is the same as the DI process for atomic line profiles. The DI reconstruction process for TiO-bands is identical to that for atomic line profiles, with one notable distinction: in the TiO-band reconstruction process, the observed spectrum is directly matched with synthetic spectra determined by atmospheric parameters, representing the photosphere and spots without the need for an additional equivalent width correction.

\subsection{Doppler Imaging of PW And}
\label{sec:di}
We performed the DI of PW And, using {\tt SpotDIPy} for Set-1, Set-2 and Set-3 separately. The surface grid was selected in trapezoidal discretization mode. Using Eq. 9.1 of \citet{Kochukhov2016}, at least 12 resolution elements along the equator is sufficient to get a correctly reconstructed image for $v\sin i=21.4$ and $R$=85\,000. However, we used 90 latitudinal zones corresponding to a total of 12,406 surface elements, for better visualization. The local line profiles are generated using the synthetic spectra that are derived from the stellar parameters provided in Table~\ref{tab:sysparams}, which were employed to represent the quiet photosphere and spots, respectively. The same procedure was also applied to TiO-band region (7000-7100~$\mbox{\AA}$). Atomic and molecular line list were extracted from VALD, while the synthetic spectra were generated using MARCS atmosphere models. We assumed the minimum spot temperature as 3800 K, which were given by \citet{Strassmeier2006}. The linear limb darkening coefficients \citep{Kostogryz2022, kostogryz2023} were determined according to the effective temperature, the surface gravity and the metallicity of PW And. The gravity darkening effect were also considered using the Eq. (31) of \citet{Espinosa2011}. However, due to the relatively low equatorial rotational velocity of PW And, the gravity darkening has no significant effect on the observed profiles.

\begin{figure}
\centering
	\includegraphics[width=\columnwidth]{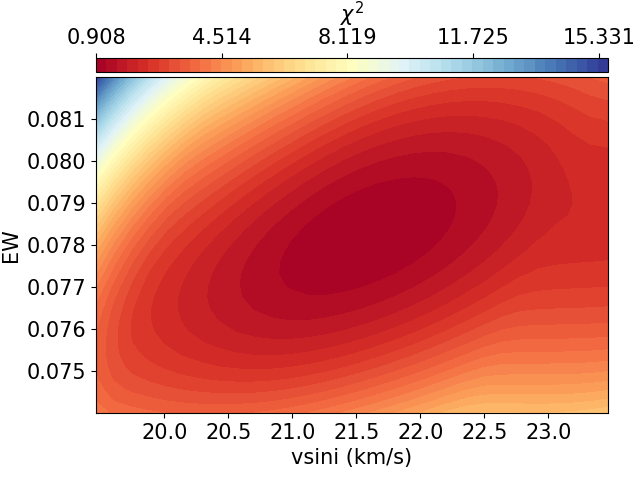}
	\includegraphics[width=\columnwidth]{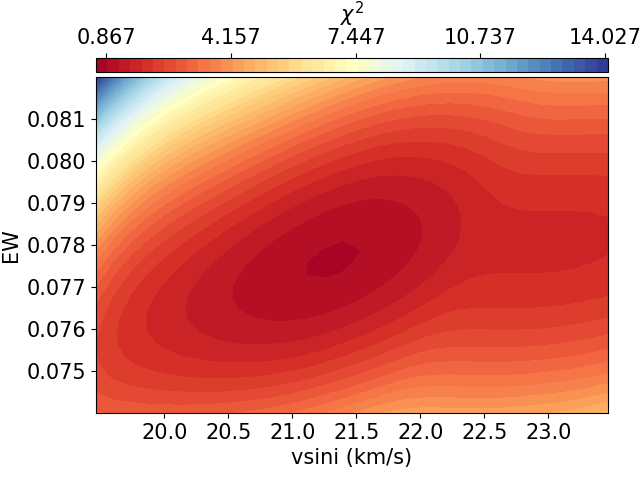}
	\includegraphics[width=\columnwidth]{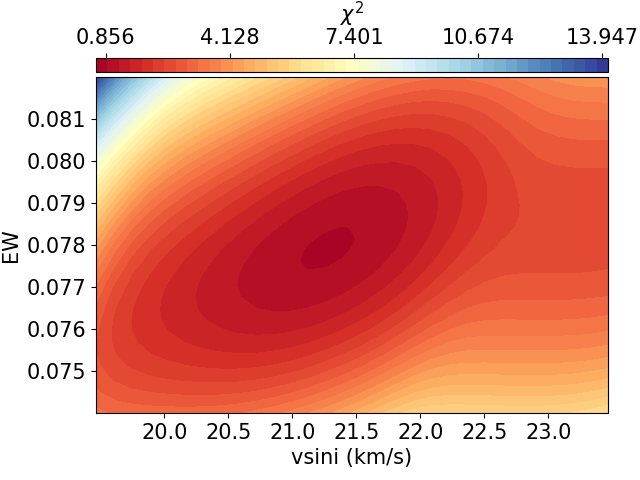}
 	\caption{2D grid search on the $EW$-$v\sin i$ plane for Set-1 (top), Set-2 (middle) and Set-3 (bottom). The colors show the $\chi^2$ values.}
    \label{fig:gridsearch}
\end{figure}

\begin{figure*}
\centering
        \includegraphics[width=0.32\textwidth]{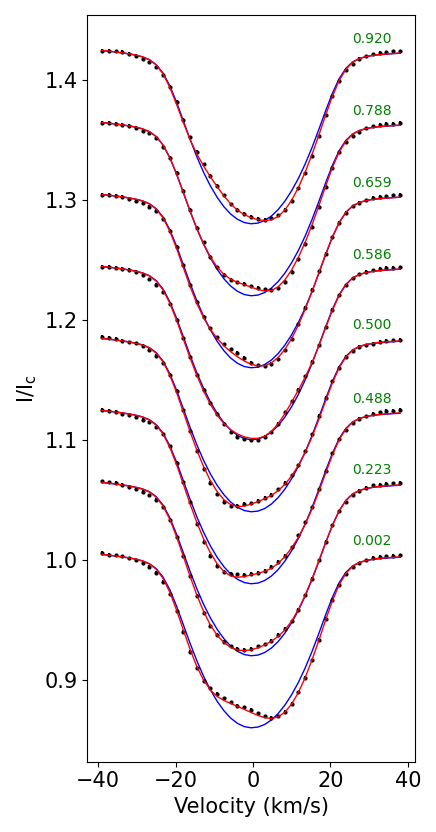}
	\includegraphics[width=0.32\textwidth]{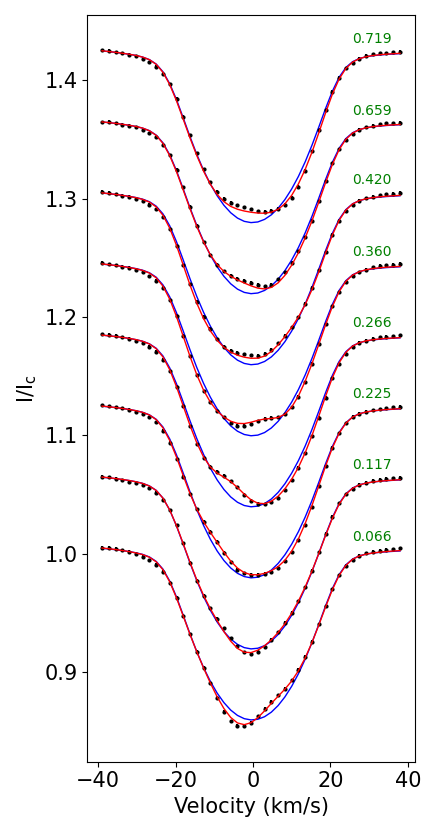}
	\includegraphics[width=0.32\textwidth]{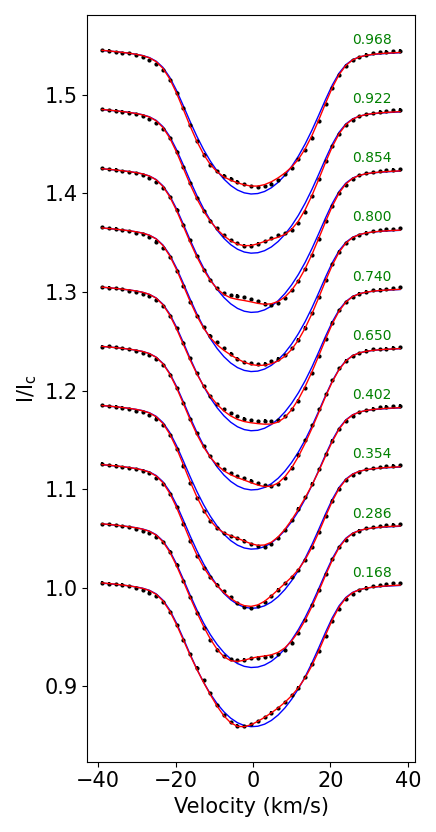}
 	\caption{Phase-ordered LSD profiles (filled circles) with the error bars derived from the observed spectra, the spotless synthetic line profiles (blue solid lines) and the best fit models (red solid lines) generated by the reconstruction process. The left, middle, and right panels are for Set-1, Set-2 and Set-3, respectively.}
    \label{fig:pwandlsd}
\end{figure*}

\begin{figure*}
\centering
        \includegraphics[width=1.0\textwidth]{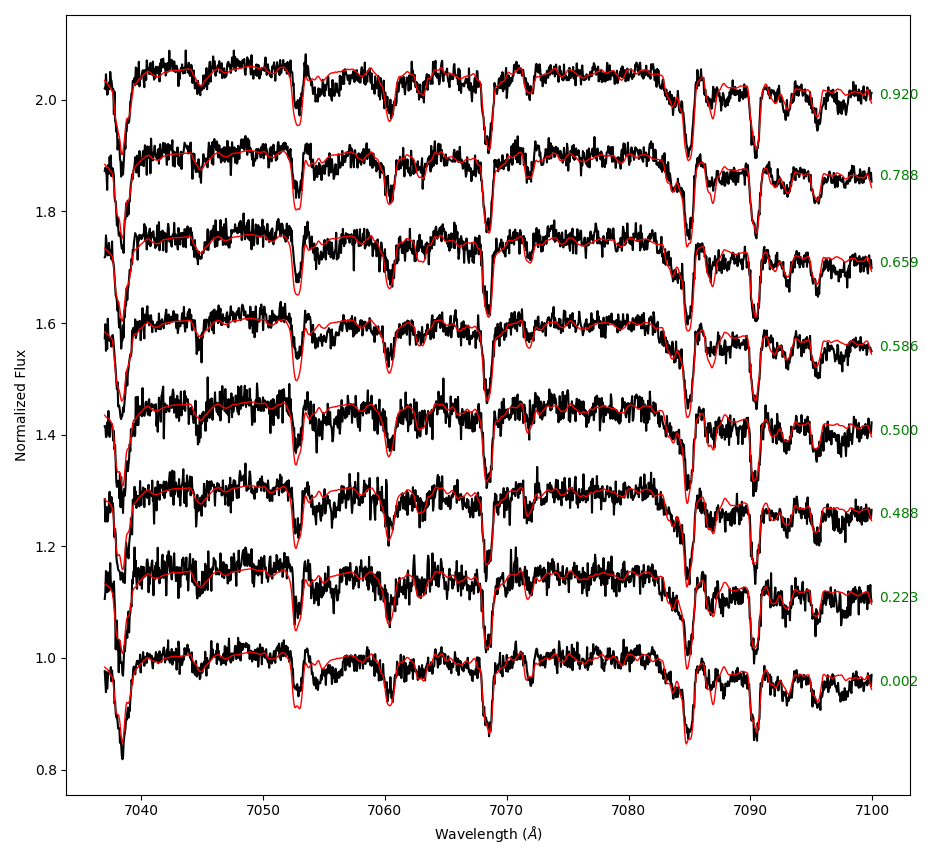}
 	\caption{Phase-ordered TiO-band profiles (black solid lines) of Set-1 and the best fit models (red solid lines) generated by the DI process.}
    \label{fig:pwandtioset1}
\end{figure*}

\begin{figure*}
\centering
	\includegraphics[width=1.0\textwidth]{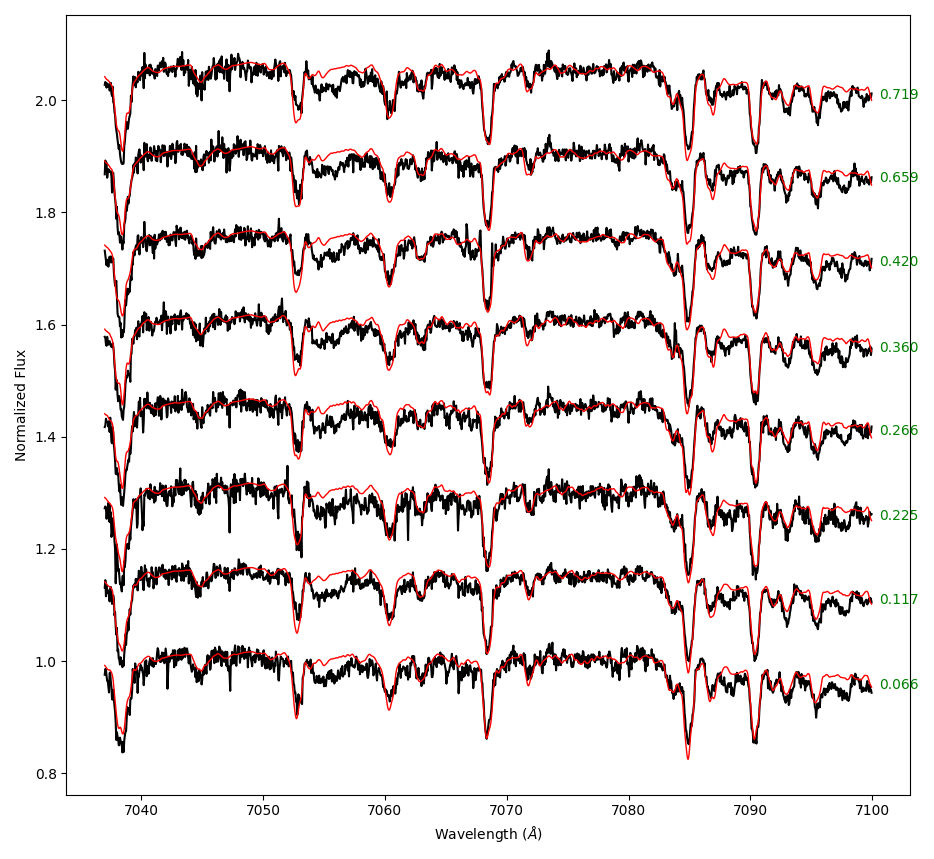}
 	\caption{Same as Figure~\ref{fig:pwandtioset1}, but for Set-2}
    \label{fig:pwandtioset2}
\end{figure*}

\begin{figure*}
\centering
	\includegraphics[width=1.0\textwidth]{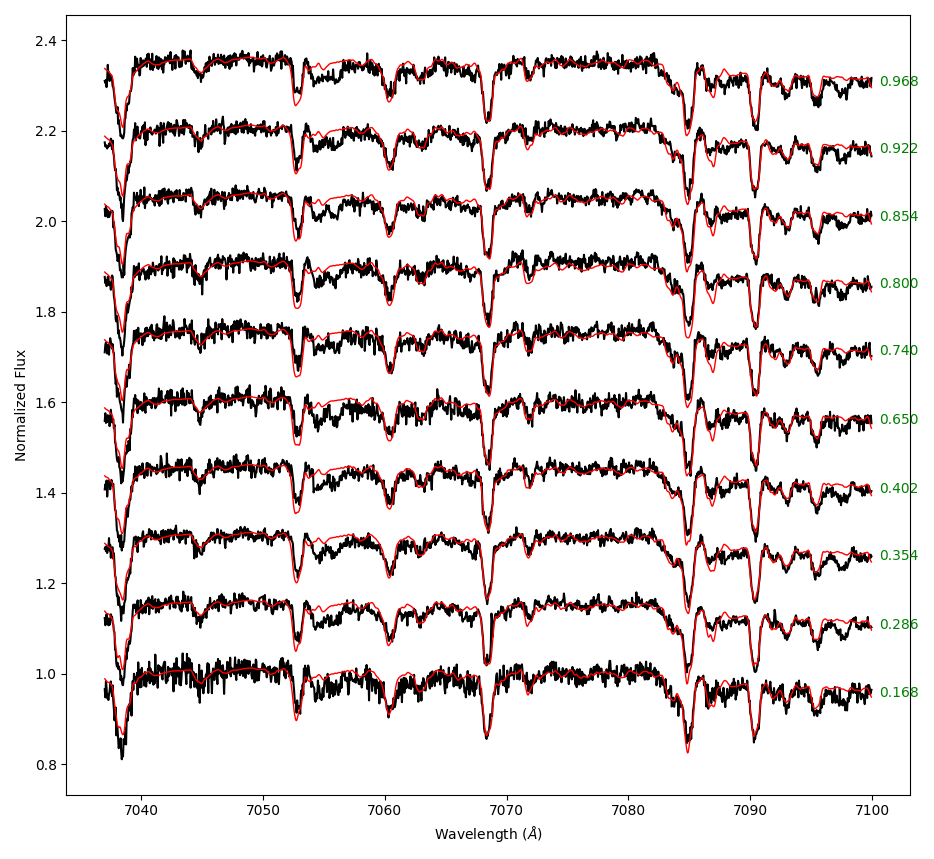}
 	\caption{Same as Figure~\ref{fig:pwandtioset1}, but for Set-3}
    \label{fig:pwandtioset3}
\end{figure*}

\begin{figure}
\centering
        \textbf{December 2015}\par\medskip
        \includegraphics[width=\columnwidth]{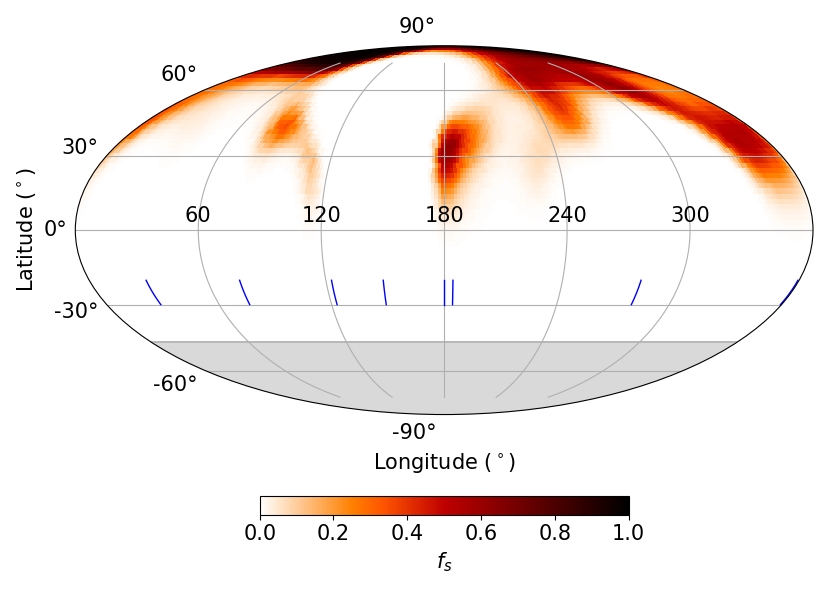}
        \textbf{September 2018}\par\medskip
	\includegraphics[width=\columnwidth]{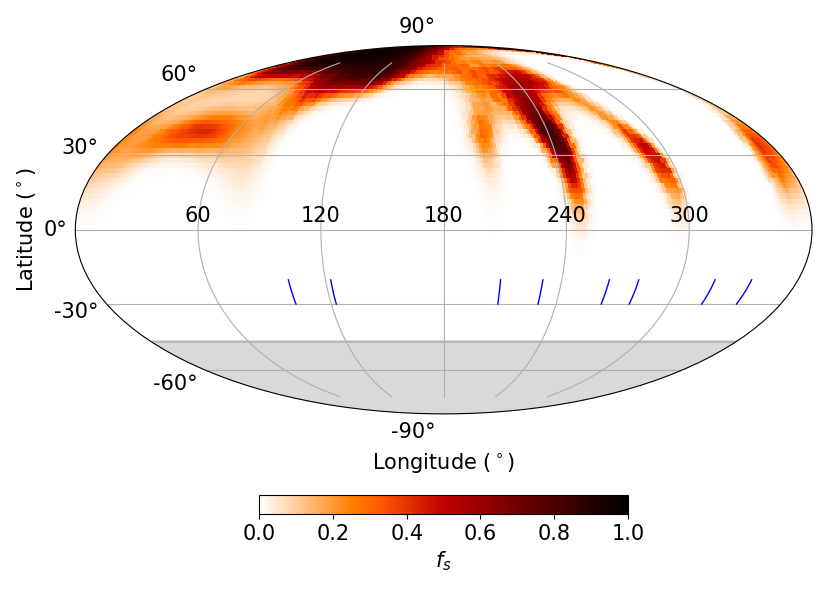}
        \textbf{December 2018}\par\medskip
	\includegraphics[width=\columnwidth]{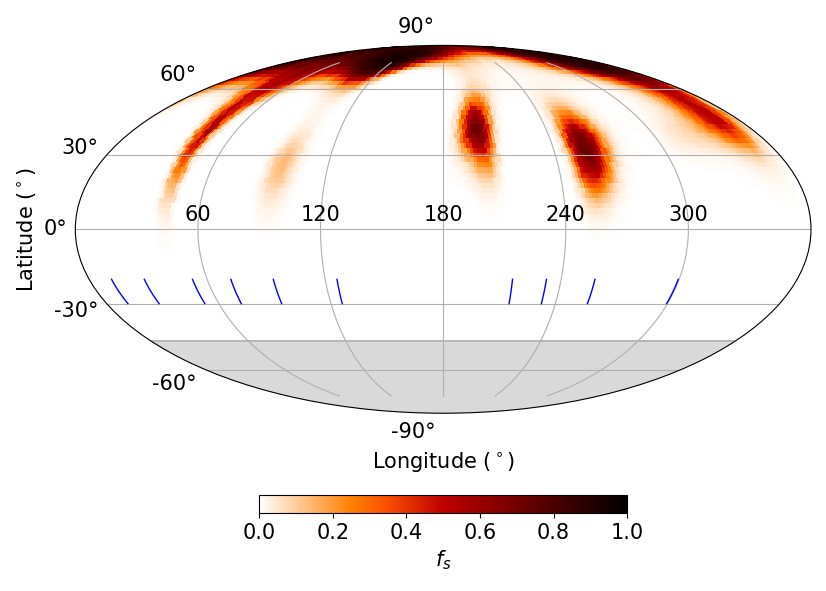}
 	\caption{Mollweide projections of the reconstructed maps of PW And in terms of spot filling factors. First three panels from top to bottom show map of Set-1, Set-2 and Set-3, respectively. The tick marks in the Mollweide projection show the phase coverage of spectral data. The shaded region around the south pole of the projection indicates unseen region of PW And's surface, due to the axial inclination.}
    \label{fig:pwandmap}
\end{figure}

It is also possible to fine-tune the stellar parameters (e.g. $v\sin i$), using $\chi^2$ minimization within {\tt SpotDIPy}. As the preliminary stellar parameters for DI, we used those given in Table~\ref{tab:sysparams}. {\tt SpotDIPy} uses an additional equivalent-width parameter (called $EW$), which affects the widths and depths of the local line profiles generated using synthetic spectra. $EW$ thus allows us to control the depth of the LSD profiles, which are also critical to circumvent DI artifacts, such as spurious polar spots \citep{Cameron1994}. To optimize $EW$, we carried out a two-dimensional grid search on the $EW$ - $v\sin i$ plane, using the $\chi^2$ minimization feature of {\tt SpotDIPy}, where the algorithm searches for minimum $\chi^2$ for a particular value of the Lagrange multiplier, for each pair of $EW$ - $v\sin i$ parameters. Contour plots of the resulting $\chi^2$ for 3 sets are shown in Figure~\ref{fig:gridsearch}. The minimum $\chi^2$ values turned out to be at $v\sin i = 21.5$$^{+1.4}_{-1.3}$~km~s$^{-1}$ for Set-1,  21.3$^{+2.2}_{-1.5}$~km~s$^{-1}$ for Set-2, 21.3$^{+1.1}_{-1.3}$~km~s$^{-1}$ for Set-3. The errors of the $v\sin i$ parameters were estimated from $\chi^2_{min}$ + 1 of $v\sin i$ values obtained from 2D grid search \citep[see][for more details]{Bevington2003}. The average $v\sin i$ was calculated as 21.4$^{+1.0}_{-0.8}$~km~s$^{-1}$, which is slightly lower than the one determined by \citet{Strassmeier2006} and almost the same as found by \citet{Llorente2021}. The adopted stellar parameters used in DI are listed in Table~\ref{tab:sysparams}. The LSD profiles, TiO-band profiles, and their respective best-fit models for all datasets are illustrated in Figure~\ref{fig:pwandlsd}, Figure~\ref{fig:pwandtioset1}-\ref{fig:pwandtioset3}, respectively. It is evident from the the Figure~\ref{fig:pwandlsd} and Figure~\ref{fig:pwandtioset1}-\ref{fig:pwandtioset3} that the models obtained through DI are highly compatible with both LSD and TiO molecular band profiles. The Mollweide projections of the resulting DI maps for Set-1, Set-2 and Set-3 are shown in Figure~\ref{fig:pwandmap}. 

Map of Set-1 shows a high-latitude spot region centered at about latitude $75^\circ$ and mid-latitude spots distributed around $30^\circ$ latitude. Set-2 and Set-3 show similar latitudinal distributions seen in the map of Set-1, with a predominant high-latitude spot and other spots spread over mid-latitudes, particularly around $30^\circ$ latitude. Although there is almost a 3-year time span between Set-1 and Set-2-3, the longitudinal distributions of starspots also show similarities, with the exception of a jump in longitude of the high-latitude spot from 2015 to 2018. The same feature is found roughly at the same longitude but with a different size when comparing Set-2 and Set-3 maps. 

The latitudinal distributions we found for all set of maps are clearly different from the surface map obtained by \citet{Strassmeier2006}, which showed dominated low-latitude spots between +40$^\circ$ and -20$^\circ$. A similar result was obtained by \citet{Kolbin2017}, who found spots covering latitudes from 30$^\circ$ to 60$^\circ$. The surface maps obtained by \citet{Gu2010}, on the other hand, exhibited a very large high-latitude spot, which extends to intermediate latitudes, making it a much more similar pattern with those in Figure~\ref{fig:pwandmap}.

The number of rotational cycles included in the spectroscopic data is a crucial factor to consider when generating DI maps. This is especially important when accounting for short-term magnetic evolution and/or a high differential rotation rate, as it can influence the presence of artificial features in the resulting maps. In the study conducted by \citet{Gu2010}, the maps were reconstructed using spectroscopic data that covered approximately 13.7 and 8.6 rotational cycles for subset 1 and subset 2, respectively. It is essential to acknowledge that reconstructed maps may exhibit artifacts due to the inclusion of spectroscopic data with relatively high rotation cycles. On the other hand, \citet{Strassmeier2006} and \citet{Kolbin2017} utilized spectroscopic data encompassing approximately 3.4 and 4.6 rotation cycles, respectively. Considering Set-1, Set-2, and Set-3 data used in this study, these values decrease to 2.9, 2.4, and 2.3, respectively. Such lower number of rotational cycles are more conducive to DI, as they reduce the likelihood of significant artificial features being introduced into the maps.

\begin{table}
\centering
\begin{threeparttable}
\caption{Adopted stellar parameters.}
\label{tab:sysparams}
\begin{tabular}{lc} 
    \hline
    \hline
    Parameter & Value \\
    \hline		
    \hline
    $vsini~{\rm[km/s]}$ & 21.4$^{+1.0}_{-0.8}$\tnote{a} \\
    $M~[M_{\odot}]$ & 0.85 $\pm$0.05\tnote{b}  \\
    $i~{\rm [^\circ]}$ & 46.0 $\pm$7\tnote{c} \\
    $P_{\rm rot}~{\rm [day]}$ & 1.756604 $\pm$0.000015\tnote{a} \\
    $T_{0}~{\rm [HJD]}$ & 2453200.0\tnote{c} \\
    $T{\rm _{eff}}~[K]$ & 5080 $\pm$28\tnote{a} \\
    $T{\rm _{spot}}~[K]$ & 3800\tnote{c} \\
    $log~g$ & 4.40 $\pm$0.09\tnote{a} \\
    $[Fe/H]$ & -0.14 $\pm$0.02\tnote{a} \\
    $\xi~{\rm[km/s]}$ & 1.93 $\pm$0.09\tnote{a} \\
    $\zeta~{\rm[km/s]}$ & 3.25 $\pm$0.06\tnote{a} \\
    \hline
    \hline
\end{tabular}
 \begin{tablenotes}
   \item[a] This study. 
   \item[b] \citet{Folsom2016}.
   \item[c] \citet{Strassmeier2006}.
 \end{tablenotes}
\end{threeparttable}
\end{table}

\subsection{Chromospheric Activity Indicators} \label{sec:chro}

The presence of chromospheric activity on PW And is known since the work by \citet{Bidelman1985}, who confirmed the existence of moderate emission in Ca~{\sc ii} H\&K lines. One of the most detailed chromospheric activity investigation of PW And was carried out by \citet{Lopez2003}, who used all chromospheric indicator lines in the optical range by employing the spectral subtraction technique and mentioned that Balmer lines are clearly visible in emission in the subtracted spectra. Depending on the ratio of $EW$(H$_{\alpha}$) and $EW$(H$_{\beta}$) \citet{Lopez2003} argued that the emission of these lines could arise from prominence-like material, whereas the ratio of the excess emission of two Ca~{\sc ii} lines (8498 \r{A} and 8542 \r{A}) arise from plage-like regions. Employing the spectral subtraction technique to H$_{\alpha}$ and Ca~{\sc ii} IRT lines, \citet{Zhang2015} obtained similar findings concerning the strong magnetic activity of PW And.  

\begin{figure}
\centering
	\includegraphics[width=\columnwidth]{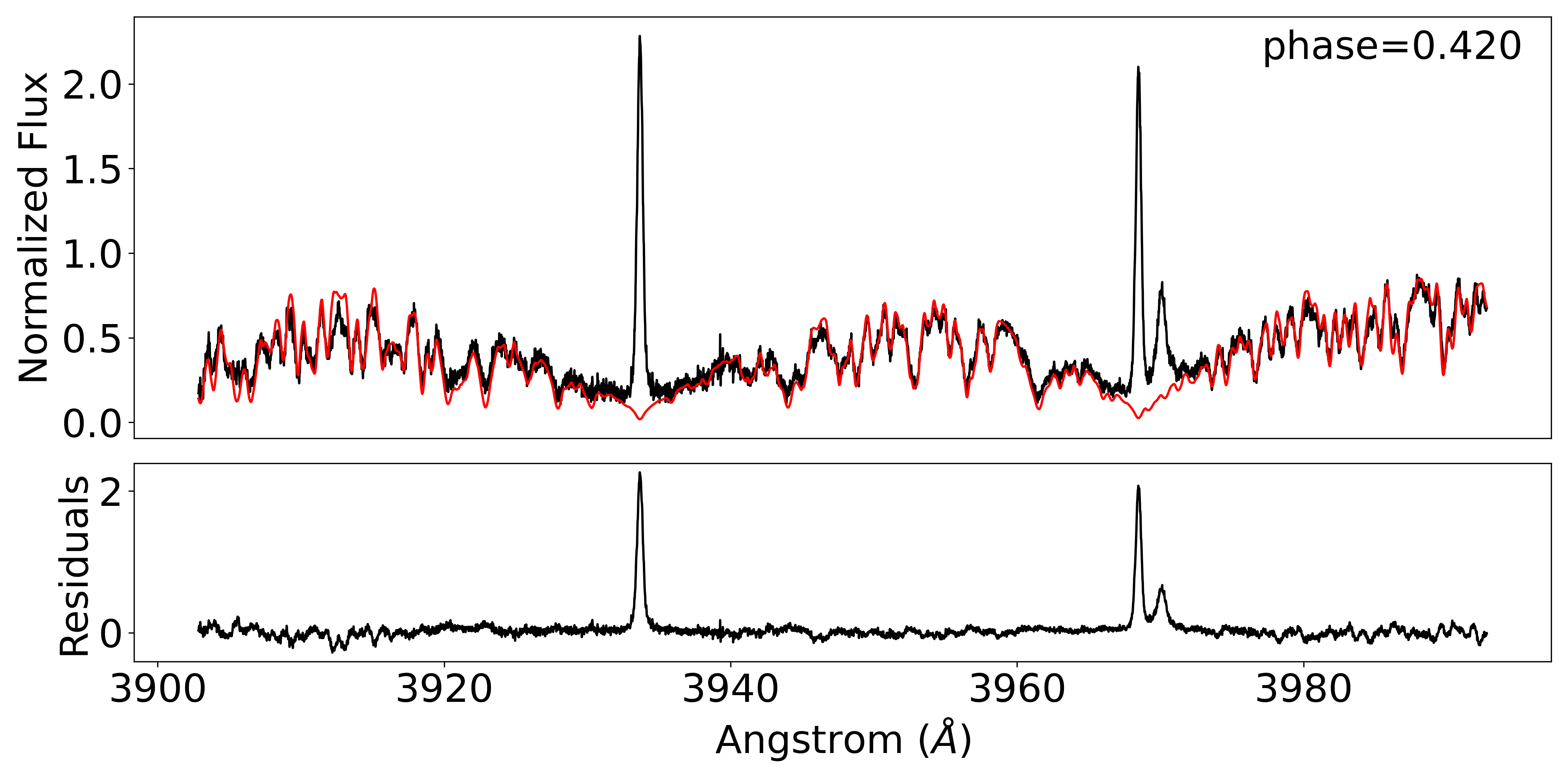}
 	\includegraphics[width=\columnwidth]{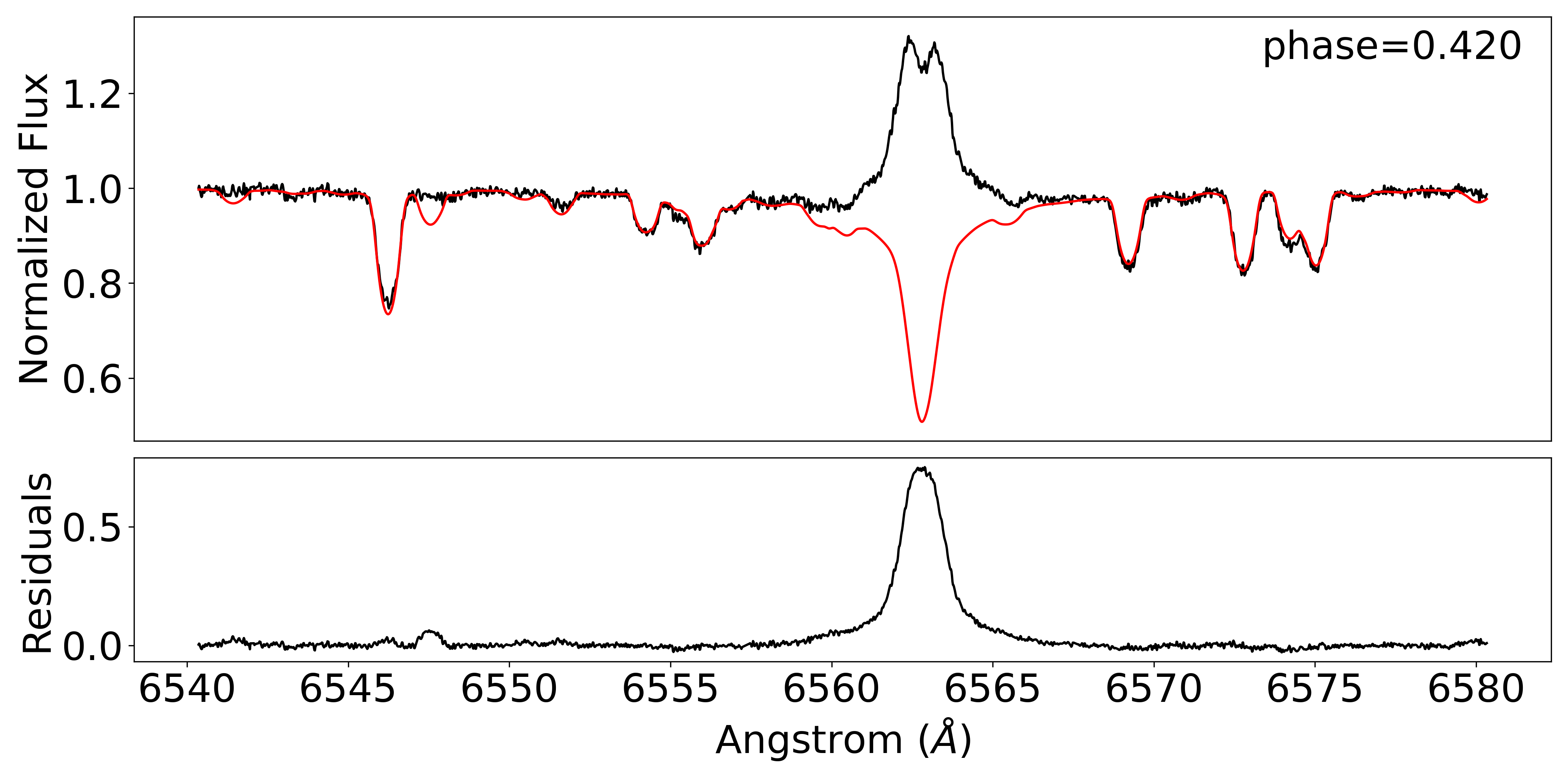}
   	\includegraphics[width=\columnwidth]{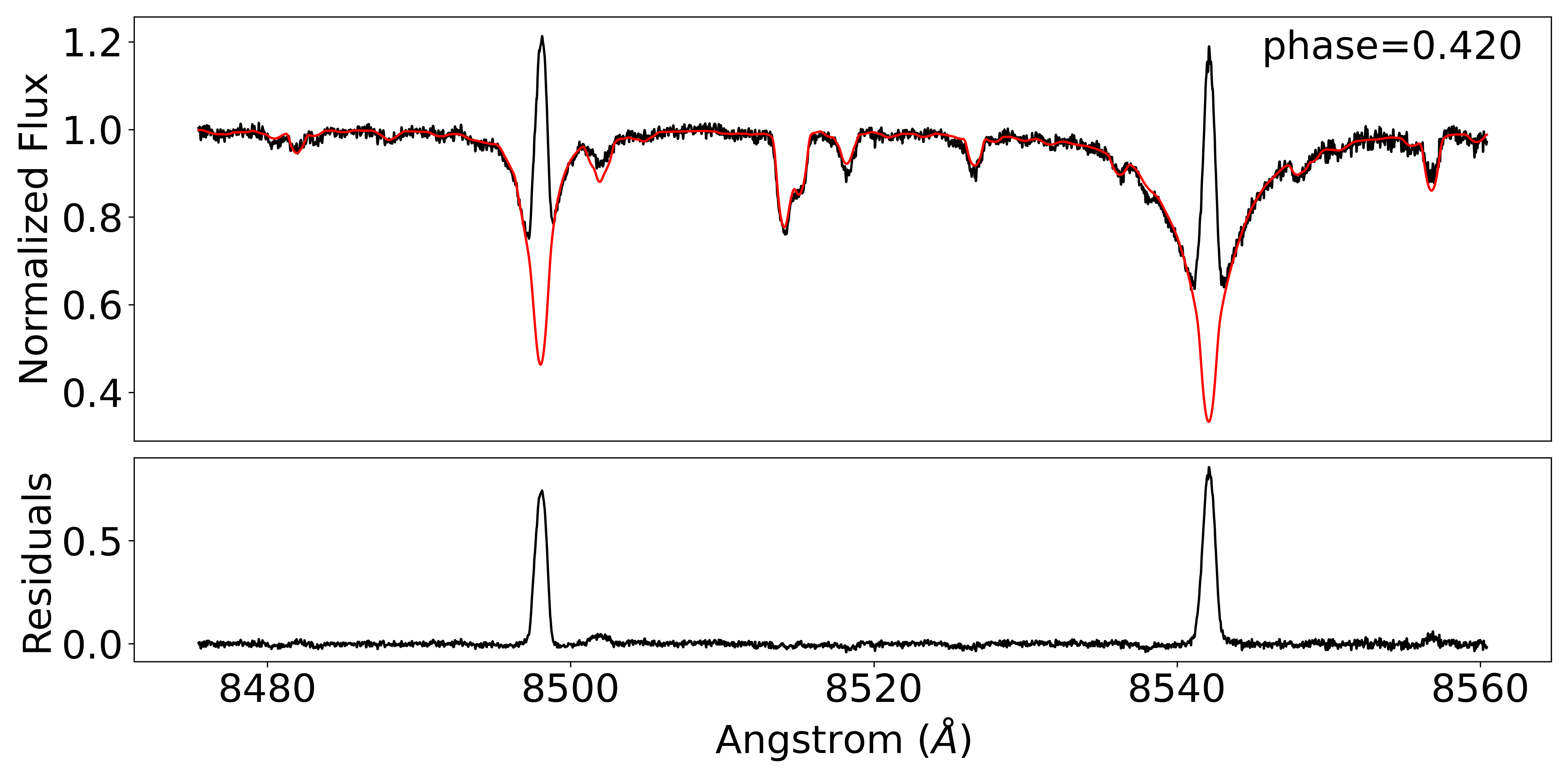}
        \includegraphics[width=\columnwidth]{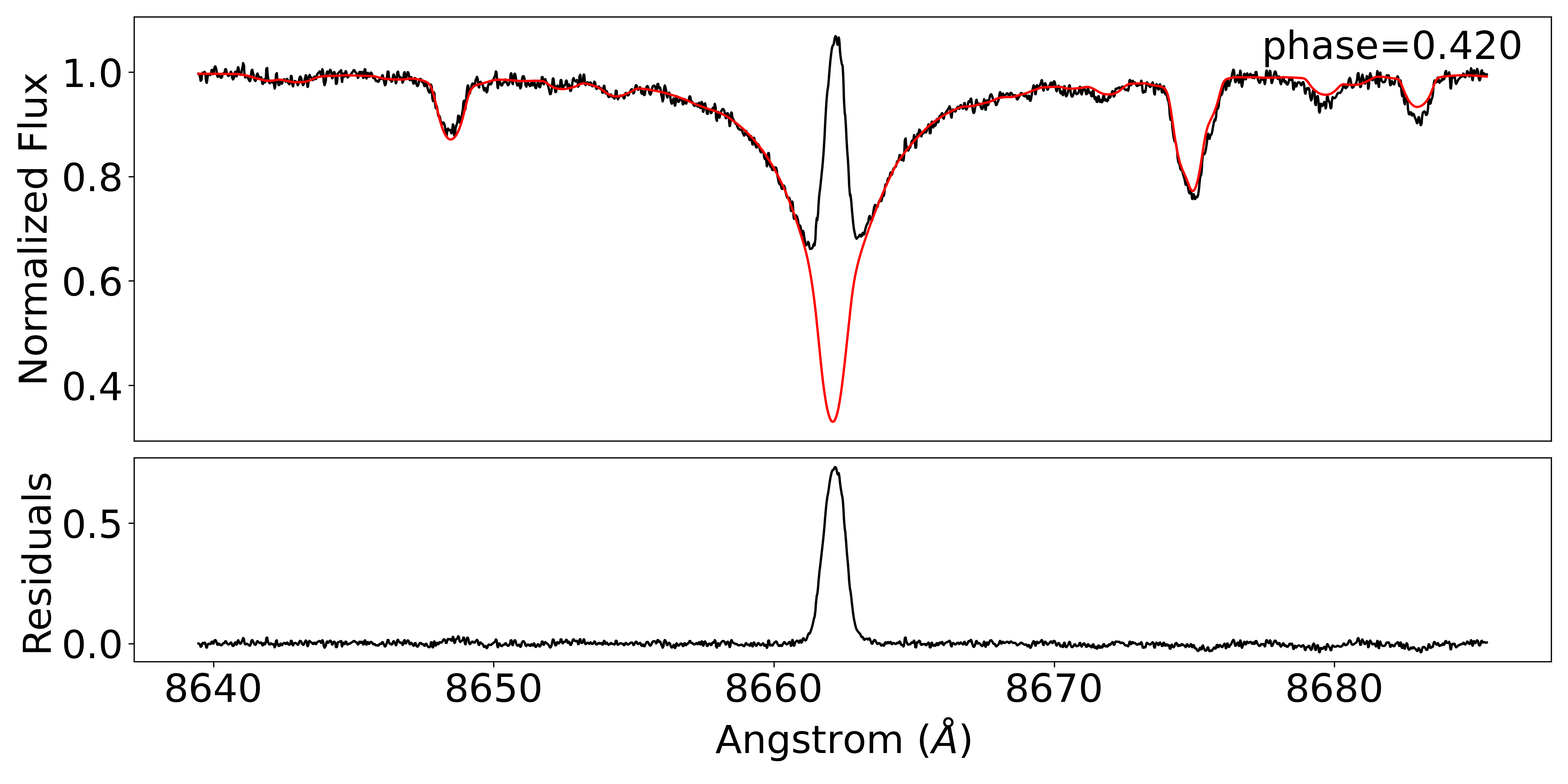}
 	\caption{An example of spectral subtraction of Ca~{\sc ii} H\&K, H$_{\alpha}$ and Ca~{\sc ii} IRT lines at the rotational phase 0.413 in the Set-2. The red solid lines show the synthesized spectrum of HD~166620. The black solid lines above and below show the observed and subtracted spectra in each panel, respectively.}
    \label{fig:ssallsample}
\end{figure}

\begin{figure}
\centering
\vspace{+0.5cm}
	\includegraphics[width=\columnwidth]{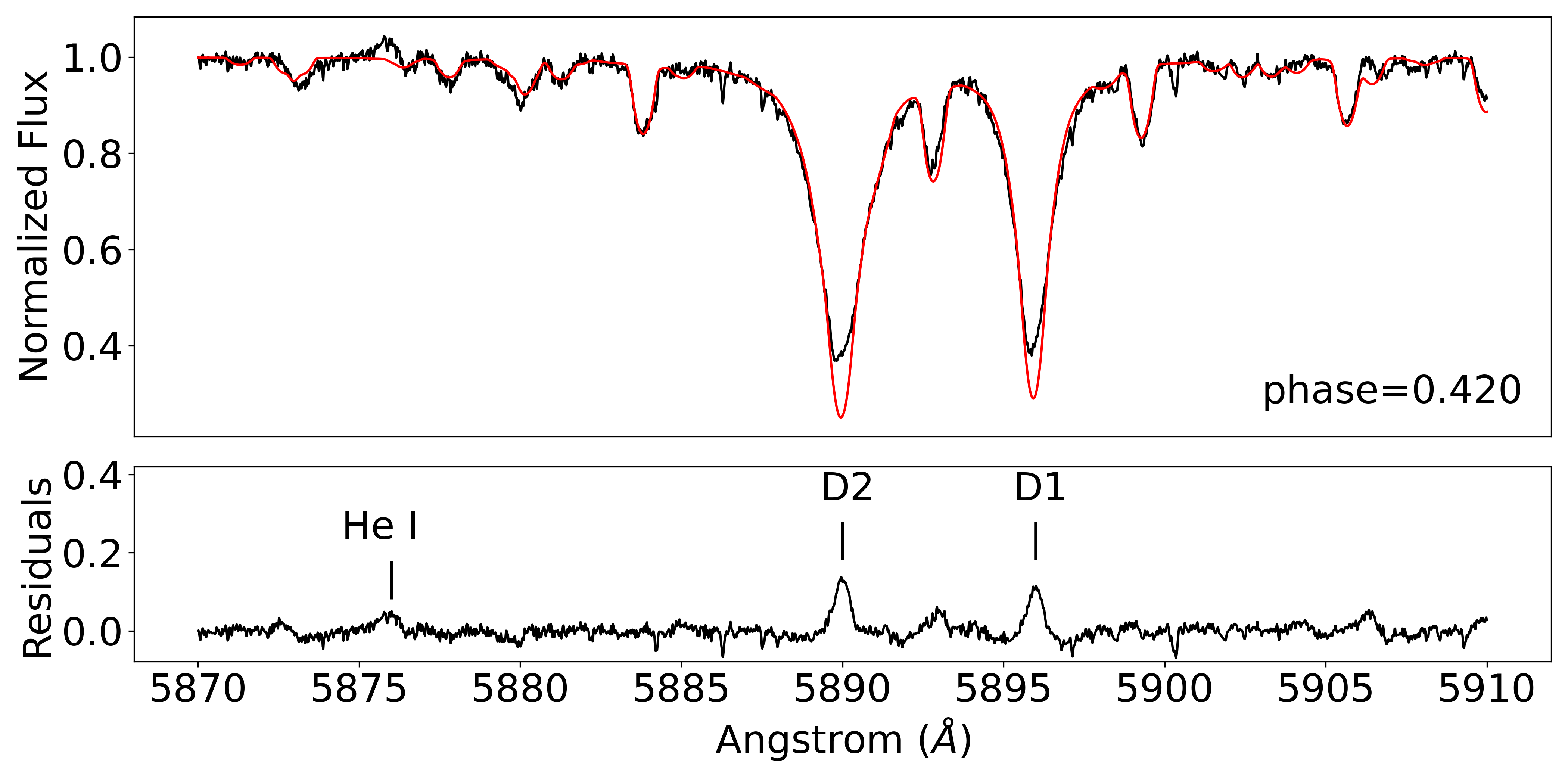}
 	\caption{Excess emissions of He~{\sc i} D$_{\rm 3}$, Na~{\sc i} (D$_{\rm 1}$, D$_{\rm 2}$) lines at phase 0.413 of Set-2.}
    \label{fig:flare}
\end{figure}

Following the chromospheric activity results in the literature, we investigate the rotational modulation of PW And's chromospheric excess emission, by applying the same method, spectral subtraction, for Ca~{\sc ii} H\&K, H$_{\alpha}$, and Ca~{\sc ii} IRT lines. By `excess emission', we hereafter mean the line emission left when the adapted synthetic spectrum obtained using the same stellar parameters of PW And was subtracted. We artificially broadened its spectral lines to the $v\sin i$ value of PW And, 21.4 km~s$^{-1}$. We then subtracted the resulting continuum-normalized spectrum from each PW And spectrum. Comparisons of PW And Ca~{\sc ii} H\&K , H$_{\alpha}$ and Ca~{\sc ii} IRT spectra with those from the synthetic spectrum are shown in Figure~\ref{fig:ssallsample}. Next, we measured the excess equivalent width (hereafter EEQW) in the Ca~{\sc ii} H\&K ($\lambda$3968 and $\lambda$3933), H$_{\alpha}$ ($\lambda$6563) and Ca~{\sc ii} IRT lines ($\lambda$8498, $\lambda$8542, and $\lambda$8662) of the subtracted spectra, by integrating over above the zero level of the subtracted profiles. We estimated uncertainties of the EEQWs, using the equation 2.3 of \citet{Schofer2021}. All EEQW measurements are listed in Table~\ref{tab:EEQWs}. The EEQW variations of Ca~{\sc ii} H\&K, H$_{\alpha}$ and Ca~{\sc ii} IRT emissions along with the rotational phase for each data-set that are shown in Figure~\ref{fig:allcomp} show similar trends among themselves. In Set-1, there is an abrupt increase in the EEQWs of all chromospheric emissions, observed during phase 0.659. A very similar phenomenon is evident for phase 0.420 of Set-2, for which the emission peaks also have distinctly higher values compared to the general trend. As a rapidly rotating K2~V star, PW~And is expected to have strong flare activity. Two such flares were detected by \citet{Lopez2003}, who were able to observe the entire pre-flare, flare and the gradual decay phases. To inspect a possible connection of abrupt increases in the EEQWs to flare activity, we measured the chromospheric flare-diagnostic lines He~{\sc i} D$_{\rm 3}$ and the Na~{\sc i} doublet. Following spectral subtraction, we found them in excess emission, as seen from Figure~\ref{fig:flare}. Therefore, we conclude that the two spectra that correspond to phases 0.659 and 0.420 of Set-1 and Set-2 data may be observed during pre-flare, flare or the gradual decay phases.

\begin{table*}
	\centering
	\caption{Measurements of EEQWs of Ca~{\sc ii} H\&K, H$\alpha$ and Ca~{\sc ii} IRT.}
	\label{tab:EEQWs}
	\begin{tabular}{ccccccc}
        \hline
        \hline
        &\multicolumn{2}{c}{Ca~{\sc ii}}& &\multicolumn{3}{c}{Ca~{\sc ii} IRT}\\
        BJD$_{Mid}$ & K & H & H$\alpha$ & $\lambda$8498 & $\lambda$8542 & $\lambda$8662\\
        \hline
        \hline
57371.33485&0.866 $\pm$0.117&1.055 $\pm$0.126&1.249 $\pm$0.071&0.595 $\pm$0.060&0.775 $\pm$0.039&0.627 $\pm$0.028\\
57372.32557&0.873 $\pm$0.101&1.050 $\pm$0.126&1.252 $\pm$0.083&0.515 $\pm$0.063&0.683 $\pm$0.038&0.563 $\pm$0.045\\
57373.31892&0.916 $\pm$0.131&0.943 $\pm$0.133&0.919 $\pm$0.063&0.458 $\pm$0.067&0.579 $\pm$0.040&0.488 $\pm$0.028\\
57375.30687&0.846 $\pm$0.062&1.023 $\pm$0.083&0.910 $\pm$0.070&0.461 $\pm$0.063&0.605 $\pm$0.035&0.493 $\pm$0.029\\
57375.45063&0.791 $\pm$0.122&1.072 $\pm$0.147&0.904 $\pm$0.075&0.481 $\pm$0.059&0.627 $\pm$0.036&0.507 $\pm$0.023\\
57376.30420&0.873 $\pm$0.085&1.115 $\pm$0.098&1.127 $\pm$0.081&0.501 $\pm$0.059&0.613 $\pm$0.049&0.524 $\pm$0.043\\
57376.32586&0.864 $\pm$0.162&1.049 $\pm$0.165&1.130 $\pm$0.078&0.478 $\pm$0.052&0.633 $\pm$0.050&0.519 $\pm$0.035\\
57376.47667&0.759 $\pm$0.062&0.991 $\pm$0.094&1.006 $\pm$0.068&0.456 $\pm$0.063&0.591 $\pm$0.038&0.491 $\pm$0.027\\
\hline
58387.36672&1.009 $\pm$0.094&1.135 $\pm$0.115&0.678 $\pm$0.069&0.399 $\pm$0.059&0.560 $\pm$0.037&0.444 $\pm$0.023\\
58387.45742&0.963 $\pm$0.086&1.248 $\pm$0.141&0.765 $\pm$0.065&0.412 $\pm$0.061&0.552 $\pm$0.033&0.442 $\pm$0.024\\
58388.40970&0.854 $\pm$0.124&1.063 $\pm$0.121&0.637 $\pm$0.060&0.410 $\pm$0.059&0.558 $\pm$0.039&0.445 $\pm$0.021\\
58388.51470&1.193 $\pm$0.104&1.479 $\pm$0.125&0.772 $\pm$0.064&0.423 $\pm$0.057&0.564 $\pm$0.033&0.456 $\pm$0.023\\
58389.40298&0.832 $\pm$0.086&1.146 $\pm$0.119&0.788 $\pm$0.066&0.423 $\pm$0.062&0.567 $\pm$0.036&0.459 $\pm$0.028\\
58389.47599&0.909 $\pm$0.096&1.142 $\pm$0.129&0.789 $\pm$0.071&0.405 $\pm$0.063&0.567 $\pm$0.030&0.465 $\pm$0.025\\
58391.39727&0.811 $\pm$0.053&1.060 $\pm$0.085&0.904 $\pm$0.062&0.442 $\pm$0.061&0.595 $\pm$0.031&0.469 $\pm$0.027\\
58391.50205&0.841 $\pm$0.054&1.057 $\pm$0.090&1.530 $\pm$0.050&0.658 $\pm$0.056&0.836 $\pm$0.030&0.706 $\pm$0.016\\
\hline
58467.44046&0.936 $\pm$0.113&1.168 $\pm$0.151&0.706 $\pm$0.071&0.446 $\pm$0.063&0.601 $\pm$0.034&0.495 $\pm$0.028\\
58468.35045&0.946 $\pm$0.159&1.134 $\pm$0.148&1.097 $\pm$0.085&0.445 $\pm$0.075&0.586 $\pm$0.046&0.471 $\pm$0.045\\
58469.35515&0.801 $\pm$0.100&1.107 $\pm$0.140&0.740 $\pm$0.072&0.415 $\pm$0.056&0.566 $\pm$0.030&0.470 $\pm$0.025\\
58469.46048&0.799 $\pm$0.077&1.025 $\pm$0.096&0.831 $\pm$0.075&0.432 $\pm$0.061&0.585 $\pm$0.039&0.472 $\pm$0.020\\
58470.31374&1.375 $\pm$0.056&1.880 $\pm$0.098&0.937 $\pm$0.070&0.438 $\pm$0.068&0.595 $\pm$0.042&0.481 $\pm$0.017\\
58470.43395&0.867 $\pm$0.069&1.135 $\pm$0.094&0.915 $\pm$0.061&0.425 $\pm$0.060&0.560 $\pm$0.044&0.467 $\pm$0.020\\
58470.51817&0.837 $\pm$0.048&1.171 $\pm$0.097&0.868 $\pm$0.062&0.419 $\pm$0.062&0.555 $\pm$0.034&0.444 $\pm$0.023\\
58471.31222&0.948 $\pm$0.158&1.064 $\pm$0.194&0.814 $\pm$0.071&0.426 $\pm$0.062&0.566 $\pm$0.035&0.469 $\pm$0.023\\
58471.43170&1.112 $\pm$0.166&1.208 $\pm$0.177&0.823 $\pm$0.072&0.401 $\pm$0.065&0.570 $\pm$0.034&0.434 $\pm$0.021\\
58471.51305&0.953 $\pm$0.162&1.055 $\pm$0.167&0.874 $\pm$0.063&0.426 $\pm$0.063&0.578 $\pm$0.036&0.466 $\pm$0.022\\

\hline
\hline
\end{tabular}
\end{table*}

\section{Conclusion and Discussion}

In this study, we analyzed high-resolution spectra of PW And considering three datasets obtained in different observing seasons, to investigate the activity nature of the star, using different methods. We also presented the most recent and accurate rotation period of PW And determined from the precise and almost continuous TESS light curves, by computing the Lomb-Scargle periodogram. The photospheric activity was investigated using the DI technique to reconstruct starspot distributions. This was achieved by simultaneously modeling both atomic line and TiO molecular band profile, utilizing a new Python-based DI code called {\tt SpotDIPy}.

\begin{figure}
\centering
        \includegraphics[width=\columnwidth]{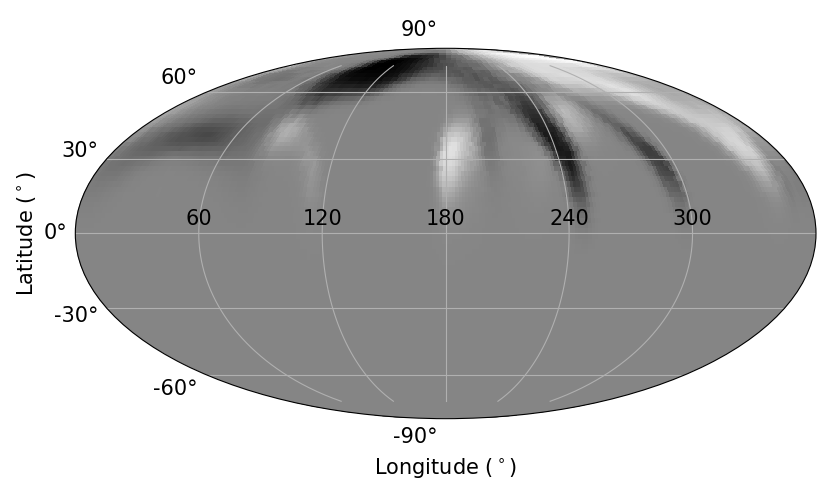}
	\includegraphics[width=\columnwidth]{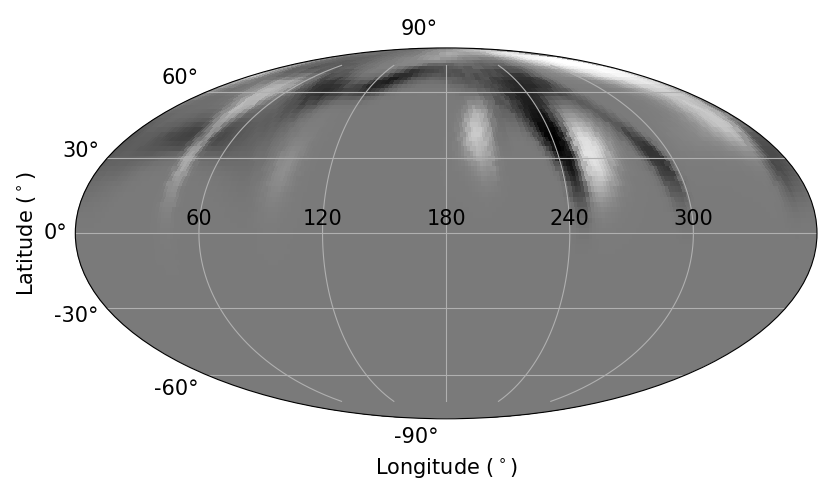}
 	\caption{Comparison of the maps reconstructed using Set-1, Set-2, and Set-3. In both panels, black indicates the predominance of spot regions from Set-2. White indicates the predominance of spot regions from Set-1 and Set-3 in upper and lower panel, respectively.}
    \label{fig:compmaps}
\end{figure}

DI maps derived from Set-1, Set-2, and Set-3 data clearly show a dominant high-latitude spot as well as mid- to low-latitude spots down to $\sim$ +30$^\circ$ latitudes. It is remarkable that the high-latitude spot continues to exist, in spite of a three-year time span between Set-1 and Set-2-3. The TESS light curves, also spanning three years, exhibit almost the same light-curve characteristics as magnetic flux continuously emerges and decays within the same phase range over the entire time span, during which one longitude region is dominant (see Figure~\ref{fig:lc_and_period}). \citet{Lopez2003} obtained the same finding using the CCF technique and mentioned that the variations of CCF bisectors found in three epochs reveal the prevalence over time of large spots or the existence of active longitudes where spots are continuously emerge. We compared the resulting surface maps of Set-1, Set-2 and Set-3 in Figure ~\ref{fig:compmaps}. It can be clearly seen that the high-latitude spot of Set-1 is located on lower longitudes than that of Set-2, while the high-latitude spots in Set-2-3 were almost centered on the same longitude. Therefore, considering (a) the explanation by \citet{Lopez2003}, (b) the TESS light curves and (c) the DI maps presented in this study, the most plausible explanation for the main component of variability in PW And is the prevalence of large spots or spot groups over time. Moreover, this might indicate that differential rotation is minimal on PW And and future observations spanning more rotation cycles could be used to place firm limits on this.

We conducted some tests to investigate the impact of S/N and phase coverage on the maps obtained for PW And. In this context, an artificial map was created with the spot distribution seen in the top panel of Figure~\ref{fig:snr_coverage_test} and artificial line profiles were generated based on the stellar parameters of PW And using this map. In the first test, we compared the maps obtained using line profiles having the same average S/N value and the phase coverage of Set-1, with the maps generated via the line profiles using the same S/N value but with a 0.05 phase interval. In the second test, unlike the first one, we compared the maps obtained from line profiles with significantly higher S/N values (Figure~\ref{fig:snr_coverage_test}). As can be seen from the resultant maps, relatively low S/N and insufficient phase coverage provide lower spot coverage than expected. Additionally, depending on the axial inclination, spots near the unseen latitudes of the stellar surface tend to have latitudes higher than what is expected. Therefore, when considering S/N and phase coverage in PW And observations, it can be said that the total spot coverage may be underestimated to some extent and the latitudes of low-latitude spots may be somewhat overestimated.

The obtained spot distributions from DI are compatible with those of \citet{Gu2010}. On the other hand, \citet{Strassmeier2006} claimed that the spots on PW And are distributed between +40$^\circ$ and -20$^\circ$ latitude without the presence of a high-latitude spot. \citet{Gu2010}, explained the reason for the large difference between the surface maps reconstructed by them and \citet{Strassmeier2006} as the lifetime of main spot structure cannot last more than one year and should change largely during its one activity cycle, similar to LQ Hya. Besides, a K0-type star’s convection zone has a larger fractional depth compared to that of a solar-mass star. When combined with the strong Coriolis effect, this results in higher emergence latitudes and strong polar magnetic fields \citep{Isik2011}, which is in accordance with the spot distributions reconstructed for PW And in this study.

\begin{figure*}
\centering
        \includegraphics[width=\textwidth]{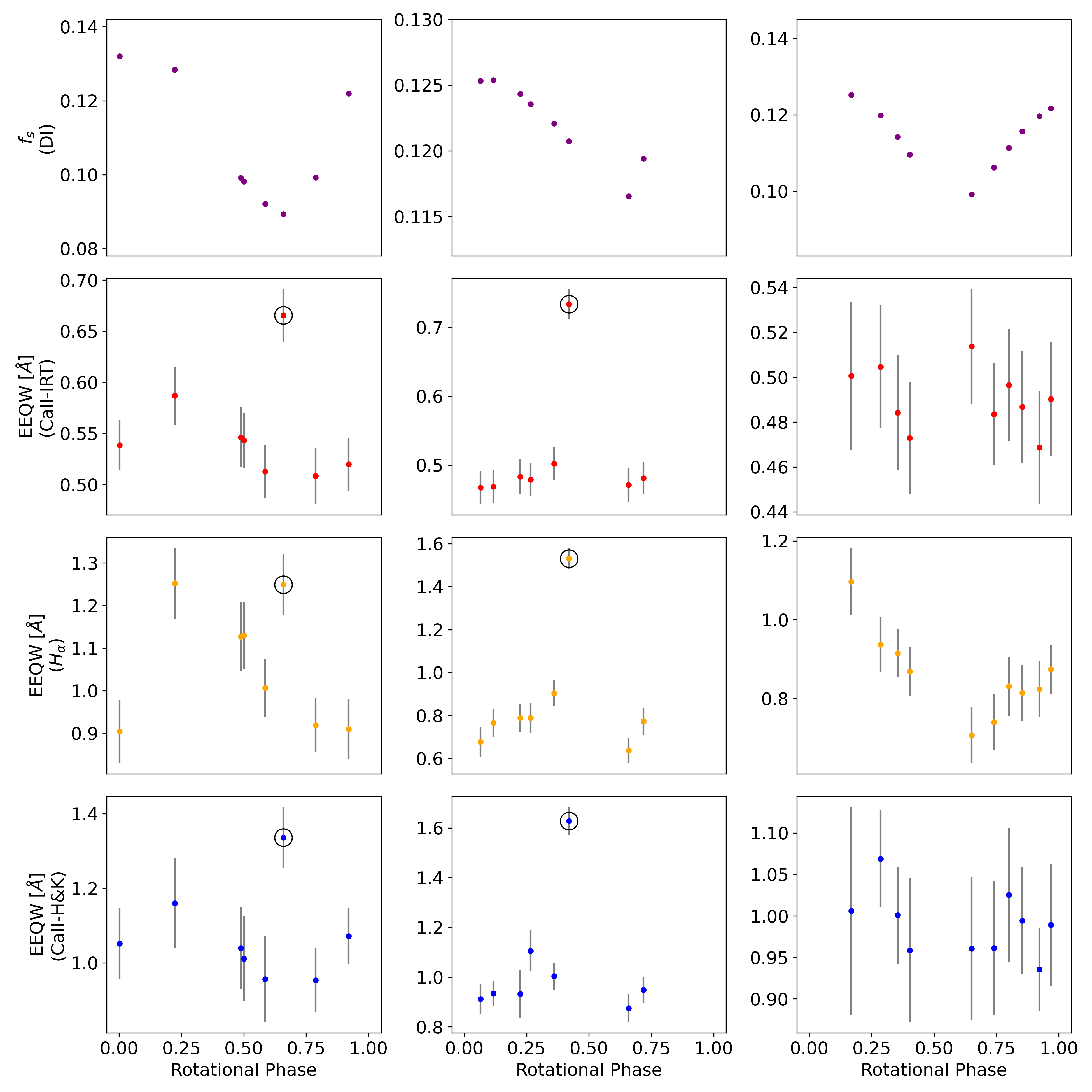}
          \caption{Comparison of photospheric and chromospheric activity induced rotational modulations. Firts row shows photospheric rotational modulations arise from cool spots obtained from DI. Third row shows EEQW variations of H$_{\alpha}$ lines, while second and last rows show averaged EEQW variatons of Ca~{\sc ii} IRT and Ca~{\sc ii} H\&K lines. respectively. From the left to the right, each column represents Set-1, Set-2 and Set-3 data, respectively. Large open circles denote flare events.}
          \label{fig:allcomp}
\end{figure*}

We show a comparison of DI-based $f_{s}$ variations along with the chromospheric diagnostics for all sets in Figure~\ref{fig:allcomp}. The EEQW variations of H$_{\alpha}$, Ca~{\sc ii} H\&K and Ca~{\sc ii} IRT emissions for the entire dataset show almost the same trend among themselves. Besides, the flare event seen during Set-1 and Set-2 is clearly traceable from the abrupt increase in EEQWs of all indicator lines. It should be noted that there is almost 2.5 hours between phases 0.360 and 0.420, where the latter corresponds to a possible flare event. \citet{Lopez2003} mentioned that there is approximately 24 hours between the flare event and the pre-flare phase during their spectral observations of PW And. Therefore, the spectrum taken during phase 0.360 most likely corresponds to the pre-flare event. 

The EEQW variation of Ca~{\sc ii} H\&K and Ca~{\sc ii} IRT lines show very similar trends to DI $f_{s}$ characteristics, while variation of EEQWs for H$\alpha$ differs slightly. Compared to the $f_{s}$ variations from DI, the chromospheric diagnostics show discrepancies due to the strong flare event in Set-2. The chromospheric excess emission variations are also in parallel with the $f_{s}$ variations obtained from DI in Set-3, although the Ca~{\sc ii} H\&K and Ca~{\sc ii} IRT are more scattered. Our findings imply that PW And's chromospheric activity patterns are spatially associated with its starspots.

Molecular bands, such as TiO-bands, offer valuable insights into the examination of cool spots on the surfaces of active stars. However, achieving precise modeling of molecular bands requires the consideration of especially the actual elemental abundances as well as the Doppler shifts to obtain a more accurate evaluation of spot coverage on stellar surfaces. In the study by \citet{Berdyugina2002}, synthetic TiO-band profiles generated from the spot distribution obtained through DI analysis of 6 atomic line profiles at 11 different rotational phases of IM Peg are seen to be in excellent agreement with the observed TiO-band profiles, proving that the recovered spot coverage from the atomic lines are reliable. In this study, a simultaneous DI process was employed to effectively model both the atomic lines (LSD profiles) and the TiO-band profiles at 7055~$\mbox{\AA}$ of PW And (see Figures~\ref{fig:pwandlsd}, ~\ref{fig:pwandtioset1}, ~\ref{fig:pwandtioset2} and ~\ref{fig:pwandtioset3}). The average S/N values obtained for atomic lines (LSDs) from Set-1, Set-2, and Set-3 are 732, 811, and 766, respectively, while for TiO-band profiles, they are 86, 118, and 100. This implies that during inversion, atomic lines have 8.5, 6.9, and 7.7 times more weight compared to molecular band profiles. Therefore, it is clear that TiO-bands exhibit significantly lower signal-to-noise ratios compared to LSD profiles, thereby giving the latter a more substantial weight on the resulting surface spot distributions in simultaneous DI processes. Nevertheless, the successful modeling of TiO-bands performed in this study validates the accuracy of the spot distributions derived from these simultaneous analyses, as also achieved by \citet{Berdyugina2002}. Therefore, in order to precisely determine the spot characteristics, the DI and TiO should be considered simultaneously.

We introduced a new Python-based DI code, {\tt SpotDIPy} that is able to reconstruct the surface brightness distribution of single stars based on two temperature approximation. {\tt SpotDIPy} was tested with the help of some simulations in comparison with the well-known DI code {\tt DoTS}. The test results showed that {\tt SpotDIPy} gave almost identical maps as those obtained via {\tt DoTS} in terms of reconstructing surface brightness distribution of stars, indicating the success of the code {\tt SpotDIPy}. SpotDIPy is an open source code that allows other users to modify and develop it freely. It is user-friendly with certain plotting GUI capabilities and is simple to use. In the near future we plan to adopt {\tt SpotDIPy} for binary stars and also to provide a surface reconstruction option for the temperature distribution of stars. The code will be accessible on GitHub$\footnote{https://github.com/EnginBahar/SpotDIPy}$ by the end of 2023.

\acknowledgments
We would like to thank the referee, Dr. Klaus Strassmeier, whose critical comments led to substantial improvement of the manuscript. HV\c{S} acknowledges the support by The Scientific And Technological Research Council Of Türkiye (T\"{U}B\.{I}TAK) through the project 1001 - 115F033. DM acknowledges financial support from the Agencia Estatal de Investigación of the Ministerio de Ciencia, Innovación through project PID2019-109522GB-C5[4]/AEI/10. OK acknowledges support by the Swedish Research Council (project 2019-03548). 

\software{DoTS \citep{Cameron1992}, SpotDIPy \citep{Bahar2023}, ExoTiC-LD \citep{grant2022}, PHOEBE2 \citep{Prsa2016}, SciPy \citep{Virtanen2020}, PyAstronomy \citep{pya2019}, Numpy \citep{Harris2020}, Autograd \citep{autograd}, Kneebow \citep{kneebow}, Astropy \citep{astropy2013, astropy2018, astropy2022}, PyQt \citep{pyqt_docu}, Matplotlib \citep{Hunter2007}, Mayavi \citep{ramachandran2011}}

\bibliography{main}

\begin{thebibliography}{}
\expandafter\ifx\csname natexlab\endcsname\relax\def\natexlab#1{#1}\fi
\providecommand{\url}[1]{\href{#1}{#1}}

\bibitem[{{Astropy Collaboration} {et~al.}(2013){Astropy Collaboration},
  {Robitaille}, {Tollerud}, {Greenfield}, {Droettboom}, {Bray}, {Aldcroft},
  {Davis}, {Ginsburg}, {Price-Whelan}, {Kerzendorf}, {Conley}, {Crighton},
  {Barbary}, {Muna}, {Ferguson}, {Grollier}, {Parikh}, {Nair}, {Unther},
  {Deil}, {Woillez}, {Conseil}, {Kramer}, {Turner}, {Singer}, {Fox}, {Weaver},
  {Zabalza}, {Edwards}, {Azalee Bostroem}, {Burke}, {Casey}, {Crawford},
  {Dencheva}, {Ely}, {Jenness}, {Labrie}, {Lim}, {Pierfederici}, {Pontzen},
  {Ptak}, {Refsdal}, {Servillat}, \& {Streicher}}]{astropy2013}
{Astropy Collaboration}, {Robitaille}, T.~P., {Tollerud}, E.~J., {et~al.} 2013,
  \aap, 558, A33

\bibitem[{{Astropy Collaboration} {et~al.}(2018){Astropy Collaboration},
  {Price-Whelan}, {Sip{\H{o}}cz}, {G{\"u}nther}, {Lim}, {Crawford}, {Conseil},
  {Shupe}, {Craig}, {Dencheva}, {Ginsburg}, {Vand erPlas}, {Bradley},
  {P{\'e}rez-Su{\'a}rez}, {de Val-Borro}, {Aldcroft}, {Cruz}, {Robitaille},
  {Tollerud}, {Ardelean}, {Babej}, {Bach}, {Bachetti}, {Bakanov}, {Bamford},
  {Barentsen}, {Barmby}, {Baumbach}, {Berry}, {Biscani}, {Boquien}, {Bostroem},
  {Bouma}, {Brammer}, {Bray}, {Breytenbach}, {Buddelmeijer}, {Burke},
  {Calderone}, {Cano Rodr{\'\i}guez}, {Cara}, {Cardoso}, {Cheedella}, {Copin},
  {Corrales}, {Crichton}, {D'Avella}, {Deil}, {Depagne}, {Dietrich}, {Donath},
  {Droettboom}, {Earl}, {Erben}, {Fabbro}, {Ferreira}, {Finethy}, {Fox},
  {Garrison}, {Gibbons}, {Goldstein}, {Gommers}, {Greco}, {Greenfield},
  {Groener}, {Grollier}, {Hagen}, {Hirst}, {Homeier}, {Horton}, {Hosseinzadeh},
  {Hu}, {Hunkeler}, {Ivezi{\'c}}, {Jain}, {Jenness}, {Kanarek}, {Kendrew},
  {Kern}, {Kerzendorf}, {Khvalko}, {King}, {Kirkby}, {Kulkarni}, {Kumar},
  {Lee}, {Lenz}, {Littlefair}, {Ma}, {Macleod}, {Mastropietro}, {McCully},
  {Montagnac}, {Morris}, {Mueller}, {Mumford}, {Muna}, {Murphy}, {Nelson},
  {Nguyen}, {Ninan}, {N{\"o}the}, {Ogaz}, {Oh}, {Parejko}, {Parley}, {Pascual},
  {Patil}, {Patil}, {Plunkett}, {Prochaska}, {Rastogi}, {Reddy Janga},
  {Sabater}, {Sakurikar}, {Seifert}, {Sherbert}, {Sherwood-Taylor}, {Shih},
  {Sick}, {Silbiger}, {Singanamalla}, {Singer}, {Sladen}, {Sooley},
  {Sornarajah}, {Streicher}, {Teuben}, {Thomas}, {Tremblay}, {Turner},
  {Terr{\'o}n}, {van Kerkwijk}, {de la Vega}, {Watkins}, {Weaver}, {Whitmore},
  {Woillez}, {Zabalza}, \& {Astropy Contributors}}]{astropy2018}
{Astropy Collaboration}, {Price-Whelan}, A.~M., {Sip{\H{o}}cz}, B.~M., {et~al.}
  2018, \aj, 156, 123

\bibitem[{{Astropy Collaboration} {et~al.}(2022){Astropy Collaboration},
  {Price-Whelan}, {Lim}, {Earl}, {Starkman}, {Bradley}, {Shupe}, {Patil},
  {Corrales}, {Brasseur}, {N{"o}the}, {Donath}, {Tollerud}, {Morris},
  {Ginsburg}, {Vaher}, {Weaver}, {Tocknell}, {Jamieson}, {van Kerkwijk},
  {Robitaille}, {Merry}, {Bachetti}, {G{"u}nther}, {Aldcroft},
  {Alvarado-Montes}, {Archibald}, {B{'o}di}, {Bapat}, {Barentsen}, {Baz{'a}n},
  {Biswas}, {Boquien}, {Burke}, {Cara}, {Cara}, {Conroy}, {Conseil}, {Craig},
  {Cross}, {Cruz}, {D'Eugenio}, {Dencheva}, {Devillepoix}, {Dietrich},
  {Eigenbrot}, {Erben}, {Ferreira}, {Foreman-Mackey}, {Fox}, {Freij}, {Garg},
  {Geda}, {Glattly}, {Gondhalekar}, {Gordon}, {Grant}, {Greenfield}, {Groener},
  {Guest}, {Gurovich}, {Handberg}, {Hart}, {Hatfield-Dodds}, {Homeier},
  {Hosseinzadeh}, {Jenness}, {Jones}, {Joseph}, {Kalmbach}, {Karamehmetoglu},
  {Ka{l}uszy{'n}ski}, {Kelley}, {Kern}, {Kerzendorf}, {Koch}, {Kulumani},
  {Lee}, {Ly}, {Ma}, {MacBride}, {Maljaars}, {Muna}, {Murphy}, {Norman},
  {O'Steen}, {Oman}, {Pacifici}, {Pascual}, {Pascual-Granado}, {Patil},
  {Perren}, {Pickering}, {Rastogi}, {Roulston}, {Ryan}, {Rykoff}, {Sabater},
  {Sakurikar}, {Salgado}, {Sanghi}, {Saunders}, {Savchenko}, {Schwardt},
  {Seifert-Eckert}, {Shih}, {Jain}, {Shukla}, {Sick}, {Simpson},
  {Singanamalla}, {Singer}, {Singhal}, {Sinha}, {Sip{H{o}}cz}, {Spitler},
  {Stansby}, {Streicher}, {{{S}}umak}, {Swinbank}, {Taranu}, {Tewary},
  {Tremblay}, {Val-Borro}, {Van Kooten}, {Vasovi{'c}}, {Verma}, {de Miranda
  Cardoso}, {Williams}, {Wilson}, {Winkel}, {Wood-Vasey}, {Xue}, {Yoachim},
  {Zhang}, {Zonca}, \& {Astropy Project Contributors}}]{astropy2022}
{Astropy Collaboration}, {Price-Whelan}, A.~M., {Lim}, P.~L., {et~al.} 2022,
  apj, 935, 167

\bibitem[{Bahar(2023)}]{Bahar2023}
Bahar, E. 2023, SpotDIPy, vv0.0.1-alpha,  Zenodo, doi:10.5281/zenodo.8386064.
\newblock \url{https://zenodo.org/badge/latestdoi/590005896}

\bibitem[{{Berdyugina}(2002)}]{Berdyugina2002}
{Berdyugina}, S.~V. 2002, Astronomische Nachrichten, 323, 192

\bibitem[{{Bevington} \& {Robinson}(2003)}]{Bevington2003}
{Bevington}, P.~R., \& {Robinson}, D.~K. 2003, {Data reduction and error
  analysis for the physical sciences}

\bibitem[{{Bidelman}(1985)}]{Bidelman1985}
{Bidelman}, W.~P. 1985, \aj, 90, 341

\bibitem[{{Blanco-Cuaresma}(2019)}]{Blanco2019}
{Blanco-Cuaresma}, S. 2019, \mnras, 486, 2075

\bibitem[{{Blanco-Cuaresma} {et~al.}(2014){Blanco-Cuaresma}, {Soubiran},
  {Heiter}, \& {Jofr{\'e}}}]{Blanco2014}
{Blanco-Cuaresma}, S., {Soubiran}, C., {Heiter}, U., \& {Jofr{\'e}}, P. 2014,
  \aap, 569, A111

\bibitem[{Byrd {et~al.}(1995)Byrd, Lu, Nocedal, \& Zhu}]{Byrd1995}
Byrd, R.~H., Lu, P., Nocedal, J., \& Zhu, C. 1995, SIAM Journal on Scientific
  Computing, 16, 1190

\bibitem[{{{\c S}enavc{\i}} {et~al.}(2018){{\c S}enavc{\i}}, {Bahar}, {Montes},
  {Zola}, {Hussain}, {Frasca}, {I{\c s}{\i}k}, \& {Y{\"o}r{\"u}ko{\v
  g}lu}}]{Senavci2018}
{{\c S}enavc{\i}}, H.~V., {Bahar}, E., {Montes}, D., {et~al.} 2018, \mnras,
  479, 875

\bibitem[{{Chiang} {et~al.}(2005){Chiang}, {Borbat}, \& {Freed}}]{Chiang2005}
{Chiang}, Y.-W., {Borbat}, P.~P., \& {Freed}, J.~H. 2005, Journal of Magnetic
  Resonance, 177, 184

\bibitem[{{Collier Cameron}(1992)}]{Cameron1992}
{Collier Cameron}, A. 1992, in Lecture Notes in Physics, Berlin Springer
  Verlag, Vol. 397, Surface Inhomogeneities on Late-Type Stars, ed. P.~B.
  {Byrne} \& D.~J. {Mullan}, 33

\bibitem[{{Collier Cameron} \& {Unruh}(1994)}]{Cameron1994}
{Collier Cameron}, A., \& {Unruh}, Y.~C. 1994, \mnras, 269, 814

\bibitem[{{Czesla} {et~al.}(2019){Czesla}, {Schr{\"o}ter}, {Schneider},
  {Huber}, {Pfeifer}, {Andreasen}, \& {Zechmeister}}]{pya2019}
{Czesla}, S., {Schr{\"o}ter}, S., {Schneider}, C.~P., {et~al.} 2019, {PyA:
  Python astronomy-related packages}, , , ascl:1906.010

\bibitem[{{Dempsey} {et~al.}(1992){Dempsey}, {Bopp}, {Strassmeier}, {Granados},
  {Henry}, \& {Hall}}]{Dempsey1992}
{Dempsey}, R.~C., {Bopp}, B.~W., {Strassmeier}, K.~G., {et~al.} 1992, \apj,
  392, 187

\bibitem[{{Donati} {et~al.}(1997){Donati}, {Semel}, {Carter}, {Rees}, \&
  {Collier Cameron}}]{Donati1997b}
{Donati}, J.-F., {Semel}, M., {Carter}, B.~D., {Rees}, D.~E., \& {Collier
  Cameron}, A. 1997, \mnras, 291, 658

\bibitem[{{Espinosa Lara} \& {Rieutord}(2011)}]{Espinosa2011}
{Espinosa Lara}, F., \& {Rieutord}, M. 2011, \aap, 533, A43

\bibitem[{{Folsom} {et~al.}(2016){Folsom}, {Petit}, {Bouvier}, {L{\`e}bre},
  {Amard}, {Palacios}, {Morin}, {Donati}, {Jeffers}, {Marsden}, \&
  {Vidotto}}]{Folsom2016}
{Folsom}, C.~P., {Petit}, P., {Bouvier}, J., {et~al.} 2016, \mnras, 457, 580

\bibitem[{Georg(2019)}]{kneebow}
Georg, U. 2019, {kneebow}: Knee or elbow detection for curves,
  \url{https://github.com/georg-un/kneebow}, ,

\bibitem[{Grant \& Wakeford(2022)}]{grant2022}
Grant, D., \& Wakeford, H.~R. 2022, Exo-TiC/ExoTiC-LD: ExoTiC-LD v3.0.0,
  vv3.0.0,  Zenodo, doi:10.5281/zenodo.7437681.
\newblock \url{https://doi.org/10.5281/zenodo.7437681}

\bibitem[{{Gray} \& {Corbally}(1994)}]{Gray1994}
{Gray}, R.~O., \& {Corbally}, C.~J. 1994, \aj, 107, 742

\bibitem[{{Griffin}(1992)}]{Griffin1992}
{Griffin}, R.~F. 1992, The Observatory, 112, 41

\bibitem[{{Gu} {et~al.}(2010){Gu}, {Collier Cameron}, \& {Kim}}]{Gu2010}
{Gu}, S.-h., {Collier Cameron}, A., \& {Kim}, K.~M. 2010, in IAU Symposium,
  Vol. 264, Solar and Stellar Variability: Impact on Earth and Planets, ed.
  A.~G. {Kosovichev}, A.~H. {Andrei}, \& J.-P. {Rozelot}, 90--92

\bibitem[{{Gustafsson} {et~al.}(2008){Gustafsson}, {Edvardsson}, {Eriksson},
  {J{\o}rgensen}, {Nordlund}, \& {Plez}}]{Gustafsson2008}
{Gustafsson}, B., {Edvardsson}, B., {Eriksson}, K., {et~al.} 2008, \aap, 486,
  951

\bibitem[{Harris {et~al.}(2020)Harris, Millman, van~der Walt, Gommers,
  Virtanen, Cournapeau, Wieser, Taylor, Berg, Smith, Kern, Picus, Hoyer, van
  Kerkwijk, Brett, Haldane, del R{\'{i}}o, Wiebe, Peterson,
  G{\'{e}}rard-Marchant, Sheppard, Reddy, Weckesser, Abbasi, Gohlke, \&
  Oliphant}]{Harris2020}
Harris, C.~R., Millman, K.~J., van~der Walt, S.~J., {et~al.} 2020, Nature, 585,
  357.
\newblock \url{https://doi.org/10.1038/s41586-020-2649-2}

\bibitem[{Hunter(2007)}]{Hunter2007}
Hunter, J.~D. 2007, Computing in Science \& Engineering, 9, 90

\bibitem[{{I{\c{s}}{\i}k} {et~al.}(2011){I{\c{s}}{\i}k}, {Schmitt}, \&
  {Sch{\"u}ssler}}]{Isik2011}
{I{\c{s}}{\i}k}, E., {Schmitt}, D., \& {Sch{\"u}ssler}, M. 2011, \aap, 528,
  A135

\bibitem[{{Kochukhov}(2016)}]{Kochukhov2016}
{Kochukhov}, O. 2016, in Lecture Notes in Physics, Berlin Springer Verlag, Vol.
  914, Lecture Notes in Physics, Berlin Springer Verlag, ed. J.-P. {Rozelot} \&
  C.~{Neiner}, 177

\bibitem[{{Kolbin} \& {Galeev}(2017)}]{Kolbin2017}
{Kolbin}, A.~I., \& {Galeev}, A.~I. 2017, in Astronomical Society of the
  Pacific Conference Series, Vol. 510, Stars: From Collapse to Collapse, ed.
  Y.~Y. {Balega}, D.~O. {Kudryavtsev}, I.~I. {Romanyuk}, \& I.~A. {Yakunin},
  417

\bibitem[{Kostogryz {et~al.}(2023)Kostogryz, Shapiro, Witzke, Grant, Wakeford,
  Stevenson, Solanki, \& Gizon}]{kostogryz2023}
Kostogryz, N., Shapiro, A., Witzke, V., {et~al.} 2023, Research Notes of the
  AAS, 7, 39

\bibitem[{Kostogryz {et~al.}(2022)Kostogryz, Witzke, Shapiro, Solanki, Maxted,
  Kurucz, \& Gizon}]{Kostogryz2022}
Kostogryz, N., Witzke, V., Shapiro, A., {et~al.} 2022, arXiv preprint
  arXiv:2206.06641

\bibitem[{{Kupka} {et~al.}(1999){Kupka}, {Piskunov}, {Ryabchikova}, {Stempels},
  \& {Weiss}}]{Kupka1999}
{Kupka}, F., {Piskunov}, N., {Ryabchikova}, T.~A., {Stempels}, H.~C., \&
  {Weiss}, W.~W. 1999, \aaps, 138, 119

\bibitem[{{Lehtinen} {et~al.}(2016){Lehtinen}, {Jetsu}, {Hackman}, {Kajatkari},
  \& {Henry}}]{Lehtinen2016}
{Lehtinen}, J., {Jetsu}, L., {Hackman}, T., {Kajatkari}, P., \& {Henry}, G.~W.
  2016, \aap, 588, A38

\bibitem[{{Llorente de Andr{\'e}s} {et~al.}(2021){Llorente de Andr{\'e}s},
  {Chavero}, {de la Reza}, {Roca-F{\`a}brega}, \& {Cifuentes}}]{Llorente2021}
{Llorente de Andr{\'e}s}, F., {Chavero}, C., {de la Reza}, R.,
  {Roca-F{\`a}brega}, S., \& {Cifuentes}, C. 2021, \aap, 654, A137

\bibitem[{{L{\'o}pez-Santiago} {et~al.}(2003){L{\'o}pez-Santiago}, {Montes},
  {Fern{\'a}ndez-Figueroa}, \& {Ramsey}}]{Lopez2003}
{L{\'o}pez-Santiago}, J., {Montes}, D., {Fern{\'a}ndez-Figueroa}, M.~J., \&
  {Ramsey}, L.~W. 2003, \aap, 411, 489

\bibitem[{{L{\'o}pez-Santiago} {et~al.}(2010){L{\'o}pez-Santiago}, {Montes},
  {G{\'a}lvez-Ortiz}, {Crespo-Chac{\'o}n}, {Mart{\'\i}nez-Arn{\'a}iz},
  {Fern{\'a}ndez-Figueroa}, {de Castro}, \& {Cornide}}]{Lopez2010}
{L{\'o}pez-Santiago}, J., {Montes}, D., {G{\'a}lvez-Ortiz}, M.~C., {et~al.}
  2010, \aap, 514, A97

\bibitem[{Maclaurin {et~al.}(2015)Maclaurin, Duvenaud, \& Adams}]{autograd}
Maclaurin, D., Duvenaud, D., \& Adams, R.~P. 2015, in ICML 2015 AutoML
  Workshop, Vol. 238, 5

\bibitem[{{Montes} {et~al.}(2000){Montes}, {Fern{\'a}ndez-Figueroa}, {De
  Castro}, {Cornide}, {Latorre}, \& {Sanz-Forcada}}]{Montes2000}
{Montes}, D., {Fern{\'a}ndez-Figueroa}, M.~J., {De Castro}, E., {et~al.} 2000,
  \aaps, 146, 103

\bibitem[{{Montes} {et~al.}(2001{\natexlab{a}}){Montes}, {L{\'o}pez-Santiago},
  {Fern{\'a}ndez-Figueroa}, \& {G{\'a}lvez}}]{Montes2001b}
{Montes}, D., {L{\'o}pez-Santiago}, J., {Fern{\'a}ndez-Figueroa}, M.~J., \&
  {G{\'a}lvez}, M.~C. 2001{\natexlab{a}}, \aap, 379, 976

\bibitem[{{Montes} {et~al.}(2004){Montes}, {L{\'o}pez-Santiago},
  {Fern{\'a}ndez-Figueroa}, \& {Ramsey}}]{Montes2004}
{Montes}, D., {L{\'o}pez-Santiago}, J., {Fern{\'a}ndez-Figueroa}, M.~J., \&
  {Ramsey}, L.~W. 2004, in IAU Symposium, Vol. 219, Stars as Suns : Activity,
  Evolution and Planets, ed. A.~K. {Dupree} \& A.~O. {Benz}, 915

\bibitem[{{Montes} {et~al.}(2001{\natexlab{b}}){Montes}, {L{\'o}pez-Santiago},
  {G{\'a}lvez}, {Fern{\'a}ndez-Figueroa}, {De Castro}, \&
  {Cornide}}]{Montes2001a}
{Montes}, D., {L{\'o}pez-Santiago}, J., {G{\'a}lvez}, M.~C., {et~al.}
  2001{\natexlab{b}}, \mnras, 328, 45

\bibitem[{{Pr{\v{s}}a} {et~al.}(2016){Pr{\v{s}}a}, {Conroy}, {Horvat}, {Pablo},
  {Kochoska}, {Bloemen}, {Giammarco}, {Hambleton}, \& {Degroote}}]{Prsa2016}
{Pr{\v{s}}a}, A., {Conroy}, K.~E., {Horvat}, M., {et~al.} 2016, \apjs, 227, 29

\bibitem[{PyQT(2012)}]{pyqt_docu}
PyQT. 2012.
\newblock
  \url{http://www.riverbankcomputing.com/static/Docs/PyQt4/html/index.html}

\bibitem[{Ramachandran \& Varoquaux(2011)}]{ramachandran2011}
Ramachandran, P., \& Varoquaux, G. 2011, Computing in Science \& Engineering,
  13, 40

\bibitem[{{Raskin} {et~al.}(2011){Raskin}, {van Winckel}, {Hensberge},
  {Jorissen}, {Lehmann}, {Waelkens}, {Avila}, {de Cuyper}, {Degroote},
  {Dubosson}, {Dumortier}, {Fr{\'e}mat}, {Laux}, {Michaud}, {Morren}, {Perez
  Padilla}, {Pessemier}, {Prins}, {Smolders}, {van Eck}, \&
  {Winkler}}]{Raskin2011}
{Raskin}, G., {van Winckel}, H., {Hensberge}, H., {et~al.} 2011, \aap, 526, A69

\bibitem[{{Ricker} {et~al.}(2015){Ricker}, {Winn}, {Vanderspek}, {Latham},
  {Bakos}, {Bean}, {Berta-Thompson}, {Brown}, {Buchhave}, {Butler}, {Butler},
  {Chaplin}, {Charbonneau}, {Christensen-Dalsgaard}, {Clampin}, {Deming},
  {Doty}, {De Lee}, {Dressing}, {Dunham}, {Endl}, {Fressin}, {Ge}, {Henning},
  {Holman}, {Howard}, {Ida}, {Jenkins}, {Jernigan}, {Johnson}, {Kaltenegger},
  {Kawai}, {Kjeldsen}, {Laughlin}, {Levine}, {Lin}, {Lissauer}, {MacQueen},
  {Marcy}, {McCullough}, {Morton}, {Narita}, {Paegert}, {Palle}, {Pepe},
  {Pepper}, {Quirrenbach}, {Rinehart}, {Sasselov}, {Sato}, {Seager},
  {Sozzetti}, {Stassun}, {Sullivan}, {Szentgyorgyi}, {Torres}, {Udry}, \&
  {Villasenor}}]{Ricker2015}
{Ricker}, G.~R., {Winn}, J.~N., {Vanderspek}, R., {et~al.} 2015, Journal of
  Astronomical Telescopes, Instruments, and Systems, 1, 014003

\bibitem[{{Sch{\"o}fer}(2021)}]{Schofer2021}
{Sch{\"o}fer}, P. 2021, PhD thesis, Georg August University of Gottingen,
  Germany

\bibitem[{{Strassmeier} \& {Rice}(2006)}]{Strassmeier2006}
{Strassmeier}, K.~G., \& {Rice}, J.~B. 2006, \aap, 460, 751

\bibitem[{{Valenti} \& {Piskunov}(1996)}]{Valenti1996}
{Valenti}, J.~A., \& {Piskunov}, N. 1996, \aaps, 118, 595

\bibitem[{Virtanen {et~al.}(2020)Virtanen, Gommers, Oliphant, Haberland, Reddy,
  Cournapeau, Burovski, Peterson, Weckesser, Bright, {van der Walt}, Brett,
  Wilson, Millman, Mayorov, Nelson, Jones, Kern, Larson, Carey, Polat, Feng,
  Moore, {VanderPlas}, Laxalde, Perktold, Cimrman, Henriksen, Quintero, Harris,
  Archibald, Ribeiro, Pedregosa, {van Mulbregt}, \& {SciPy 1.0
  Contributors}}]{Virtanen2020}
Virtanen, P., Gommers, R., Oliphant, T.~E., {et~al.} 2020, Nature Methods, 17,
  261

\bibitem[{{Zhang} {et~al.}(2015){Zhang}, {Pi}, \& {Zhu}}]{Zhang2015}
{Zhang}, L.-Y., {Pi}, Q.-F., \& {Zhu}, Z.-Z. 2015, Research in Astronomy and
  Astrophysics, 15, 252

\bibitem[{Zhu {et~al.}(1997)Zhu, Byrd, Lu, \& Nocedal}]{Zhu1997}
Zhu, C., Byrd, R.~H., Lu, P., \& Nocedal, J. 1997, ACM Transactions on
  Mathematical Software, 23, https://doi.org/10.1145/279232.279236

\end{thebibliography}

\appendix

\section{test results}
\label{sec:appena}

\begin{figure*}[!h]
\centering
\includegraphics[width=\textwidth]{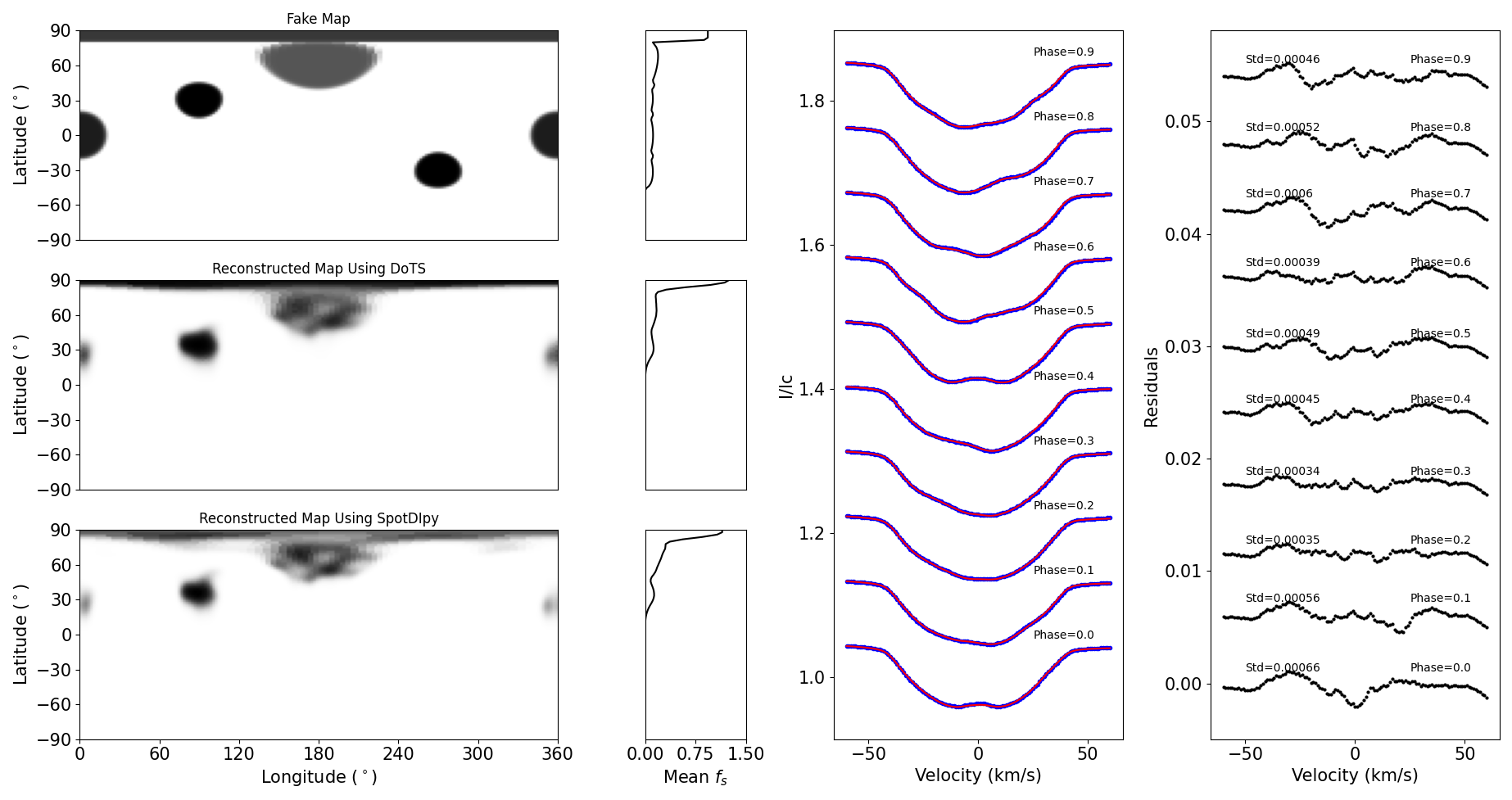}
\caption{A comparison simulation of {\tt DoTS} and {\tt SpotDIPy}. In the first column, original spotted map (top) and reconstructed maps using {\tt DoTS} (middle) and {\tt SpotDIPy} (bottom), respectively. The third column shows latitudinal cross-section of the spots recovered. The third column shows the best fit models generated by {\tt DoTS} (blue filled circles) and {\tt SpotDIPy} (red solid lines). Differences between the best fit models with their standard deviations are shown in the fourth column. Phases corresponds to each best fits and residuals are indicated. Original spotted map and artificial synthetic line profiles with 500 S/N generated by {\tt DoTS} under the 30$^{\circ}$ axial inclination.}
\label{fig:comp_dots_sdip30}
\end{figure*}

\begin{figure*}
\centering
\includegraphics[width=\textwidth]{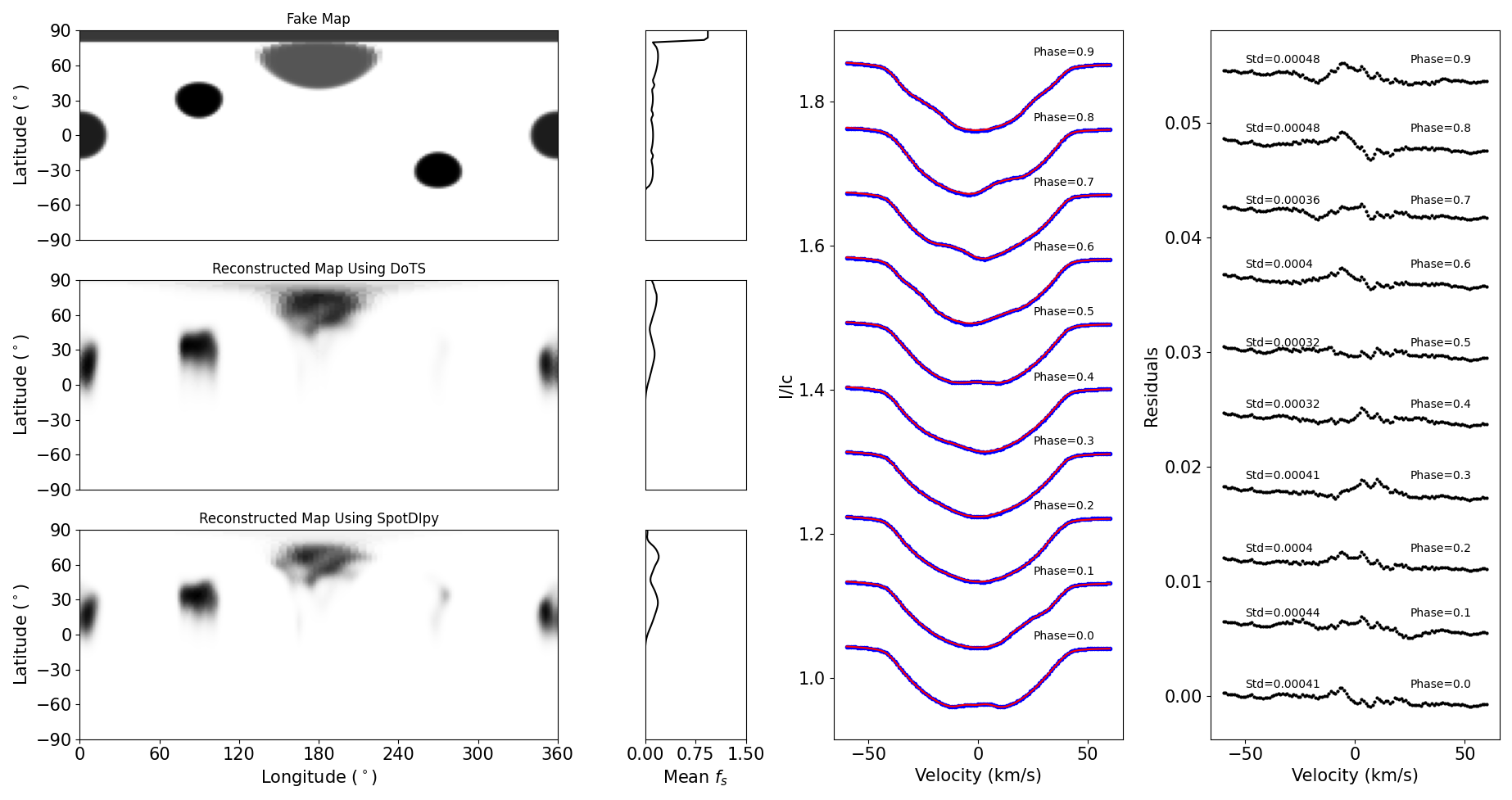}
\caption{Same as Figure~\ref{fig:comp_dots_sdip30}, but for 60$^{\circ}$ axial inclination.}
\label{fig:comp_dots_sdip60}
\end{figure*}

\begin{figure*}
\centering
\includegraphics[width=\textwidth]{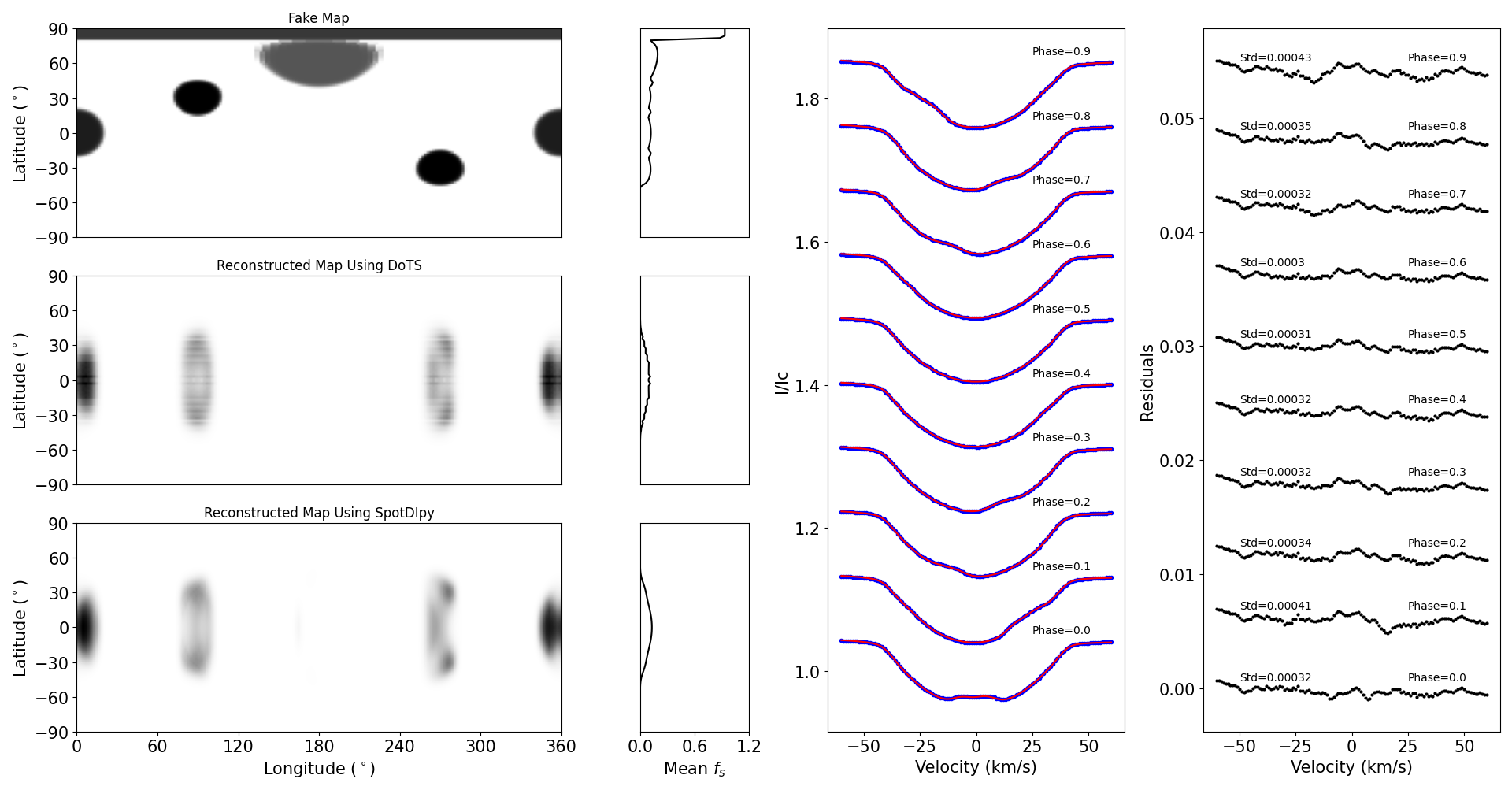}
\caption{Same as Figure~\ref{fig:comp_dots_sdip30}, but for 90$^{\circ}$ axial inclination.}
\label{fig:comp_dots_sdip90}
\end{figure*}

\begin{figure*}
\centering
\includegraphics[width=\textwidth]{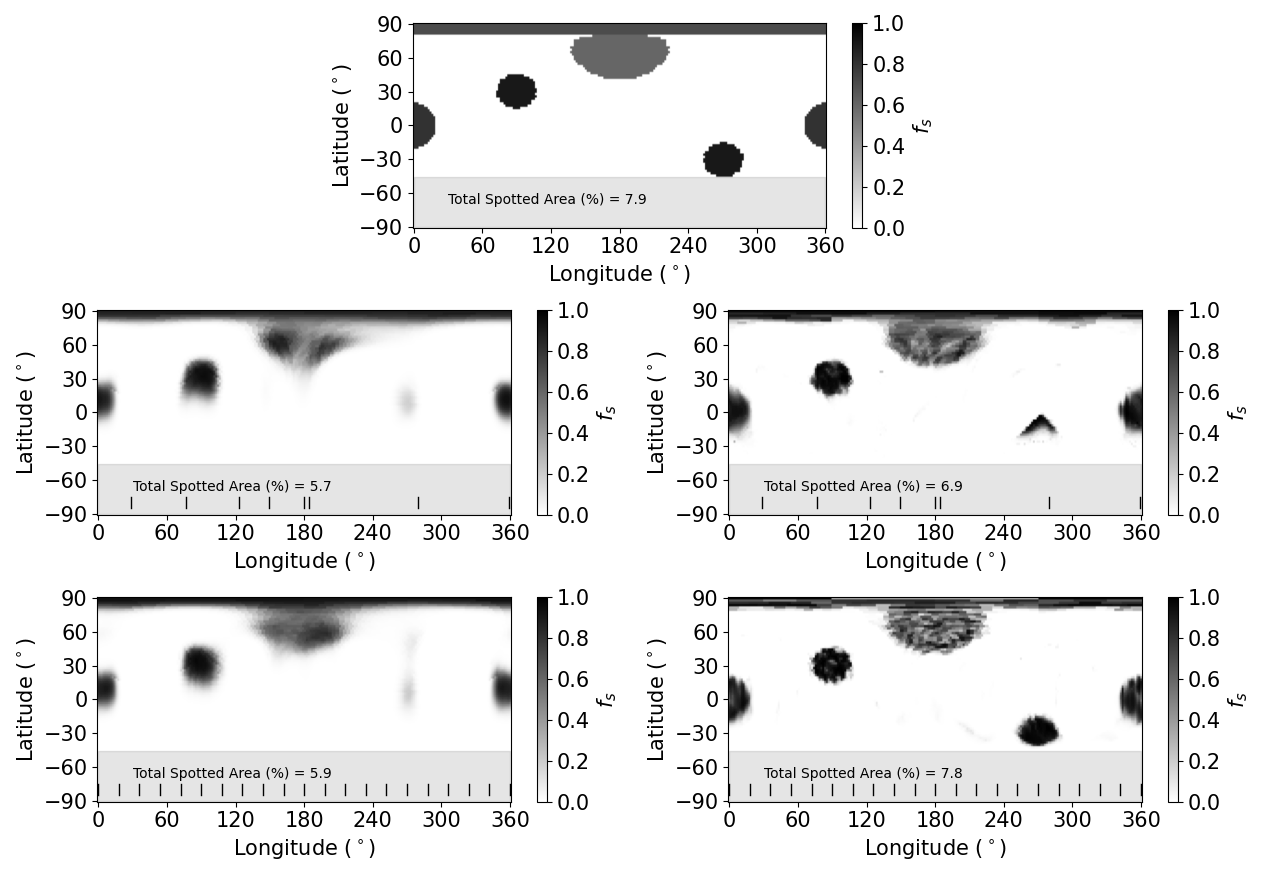}
\caption{Test results for the effect of S/N and phase coverage on the reconstructed surface images. The top map is an artificial map containing spots placed at different latitudes and longitudes, and all artificial line profiles have been generated from this map. The middle left image was reconstructed using line profiles with the same phase coverage and S/N ratio as Set-1 data, while the lower left image used the same S/N ratio but with a 0.05 phase sampling interval. In the middle right image, we maintained the same phase coverage but with a significantly higher S/N ratio, while the lower right image combined a 0.05 phase sampling interval with a very high S/N ratio. The shaded region around the south pole of the projection indicates unseen region of the stellar surface.}
\label{fig:snr_coverage_test}
\end{figure*}

\end{document}